\documentclass{aa}  
\usepackage{multirow}
\usepackage{xcolor}
\usepackage{lscape}
\usepackage{graphicx}
\usepackage{textcomp}
\usepackage{placeins}
\usepackage{lscape}
\usepackage{txfonts}
\begin{document}

\title{Large Interferometer For Exoplanets (LIFE)}
\subtitle{X. Detectability of currently known exoplanets and synergies with future IR/O/UV reflected-starlight imaging missions}

   \author{\'Oscar Carri\'on-Gonz\'alez\inst{1}\thanks{\email{oscar.carrion@obspm.fr, oscar.carrion.gonzalez@gmail.com}}
         \and Jens Kammerer\inst{2}
         \and Daniel Angerhausen\inst{3,4}
         \and Felix Dannert\inst{3}
         \and Antonio Garc\'ia Mu\~noz\inst{5}
         \and Sascha P. Quanz\inst{3,4}
         \and Olivier Absil\inst{6}
         \and Charles A. Beichman\inst{7}
         \and Julien H. Girard\inst{2}
         \and Bertrand Mennesson\inst{8}
         \and Michael R. Meyer\inst{9}
         \and Karl R. Stapelfeldt\inst{8}
         \and The LIFE Collaboration\inst{10}
          }

   \institute{
         LESIA, Observatoire de Paris, Universit\'e PSL, CNRS, Sorbonne Universit\'e, Universit\'e Paris Cit\'e, 5 place Jules Janssen, 92195 Meudon, France
         \and Space Telescope Science Institute (STScI), 3700 San Martin Dr, Baltimore, MD 21218, USA
         \and ETH Zurich, Institute for Particle Physics \& Astrophysics, Wolfgang-Pauli-Str. 27, 8093 Zurich, Switzerland
         \and National Center of Competence in Research PlanetS, Gesellschaftsstrasse 6, 3012 Bern, Switzerland
         \and Universit\'e Paris-Saclay, Universit\'e Paris Cit\'e, CEA, CNRS, AIM, 91191, Gif-sur-Yvette, France
         \and STAR Institute, Universit\'e de Li\`ege, All\'e du Six Ao\^ut 19C, 4000 Li\`ege, Belgium
         \and NASA Exoplanet Science Institute, Jet Propulsion Laboratory, California Institute of Technology, 1200 East California Blvd, Pasadena, CA 91125, USA
         \and Jet Propulsion Laboratory, California Institute of Technology, 4800 Oak Grove Drive, Pasadena, CA 91109, USA
         \and Department of Astronomy, University of Michigan, Ann Arbor, MI 48109, USA
         \and \url{http://www.life-space-mission.com/}
             }


 
  \abstract
   {The next generation of space-based observatories will characterize the atmospheres of low-mass, temperate exoplanets with the direct-imaging technique. This will be a major step forward in our understanding of exoplanet diversity and the prevalence of potentially habitable conditions beyond the Earth.} 
   {We compute a list of currently known exoplanets detectable with the mid-infrared Large Interferometer For Exoplanets (LIFE) in thermal emission. We also compute the list of known exoplanets accessible to a notional design of the future Habitable Worlds Observatory (HWO), observing in reflected starlight.}
   {With a pre-existing statistical methodology, we processed the NASA Exoplanet Archive and computed orbital realizations for each known exoplanet. We derived their mass, radius, equilibrium temperature, and planet-star angular separation. We used the \texttt{LIFEsim} simulator to compute the integration time ($t_{int}$) required to detect each planet with LIFE. A planet is considered detectable if a broadband signal-to-noise ratio $S/N$=7 is achieved over the spectral range 4$-$18.5~$\mu$m in $t_{int}\leq$100~hours. We tested whether the planet is accessible to HWO in reflected starlight based on its notional inner and outer working angles, and minimum planet-to-star contrast.}
   {LIFE’s reference configuration (four 2-m telescopes with 5\% throughput and a nulling baseline between 10--100~m) can detect 212 known exoplanets within 20~pc. Of these, 55 are also accessible to HWO in reflected starlight, offering a unique opportunity for synergies in atmospheric characterization. LIFE can also detect 32 known transiting exoplanets. Furthermore, we find 38 LIFE-detectable planets orbiting in the habitable zone, of which 13 have $M_p$<5$M_\oplus$ and eight have 5$M_\oplus$<$M_p$<10$M_\oplus$.}
   {LIFE already has enough targets to perform ground-breaking analyses of low-mass, habitable-zone exoplanets, a fraction of which will also be accessible to other instruments.} 



   \maketitle
%

\section{Introduction} \label{sec:introduction}

With more than 5000 exoplanets discovered to date, characterizing their atmospheres remains a challenge for most of the cases.
Moving from detection to atmospheric characterization is a key step towards understanding the diversity of worlds in our galaxy.
This will put our Solar System and the Earth into a broader context, allowing us to derive statistical conclusions about the prospects for habitability outside our planet.
For this, a new generation of space-borne telescopes capable of characterizing potential Earth analogues will be needed.

Transit measurements are so far the main technique available to discover and characterize new exoplanets, with multiple ground-based facilities and space missions such as CoRoT \citep{baglinetal2006}, Kepler \citep{boruckietal2010}, K2 \citep{howelletal2014}, TESS \citep{rickeretal2014}, and PLATO \citep{raueretal2014}.
The population of exoplanets observed in transit is biased towards giant planets on short-period orbits because these create stronger signals and are more likely to be observed.
The James Webb Space Telescope \citep[JWST,][]{gardneretal2006, gardneretal2023} will be able to observe and characterize spectroscopically some low-mass exoplanets with lower equilibrium temperatures, mostly orbiting nearby M dwarfs \citep{beichmanetal2014, barstow-irwin2016}.
This will reveal whether small planets around such active stars can retain an atmosphere \citep{greeneetal2023} and, if so, will probe their upper atmospheric layers \citep[e.g.][]{raueretal2011, wunderlichetal2019}.
Planned for launch in 2029, the Ariel mission \citep{tinettietal2018} will study the atmospheres of hundreds of warm and hot transiting exoplanets, including some low-mass targets.

Earth analogues orbiting Sun-like stars remain however out of reach for atmospheric characterization with these transit missions.
Temperate long-period exoplanets are less likely to be found in transit and, even if found, inherent phenomena such as light refraction will limit the potential for atmospheric characterization \citep{garciamunozetal2012, misraetal2014}.
Direct-imaging observations are thus needed to analyse the atmospheres of a significant number of temperate low-mass exoplanets.

Several direct-imaging facilities have been proposed for the next decades.
On the ground, upcoming extremely large telescopes (ELTs) will be equipped with direct-imaging instruments. The first generation of these instruments will mostly directly image exoplanets in thermal emission \citep[e.g.][]{quanzetal2015, bowensetal2021}, although instruments such as ELT/HARMONI and TMT/MODHIS might get to directly image some mature planets in reflected starlight \citep{houlleetal2021, mawetetal2022}. A next generation of ground-based direct-imaging instruments will be devoted to observing mature exoplanets in reflected starlight \citep{kasperetal2021, fitzgeraldetal2022, malesetal2022}. ELTs will generally be sensitive to exoplanets around the nearest stars, some of them potentially down to Earth-like sizes \citep{kasperetal2021}.

In space, the Nancy Grace Roman Space Telescope\footnote{Previously named Wide-Field Infrared Survey Telescope (WFIRST) } \citep[hereafter Roman,][]{spergeletal2015}, planned for launch in late 2026 or early 2027, will be equipped with a Coronagraph Instrument observing at visible wavelengths \citep{kasdinetal2020}.
This instrument will directly image cold giant planets in reflected starlight \citep{carriongonzalezetal2021a} and also measure the thermal emission of young exoplanets \citep{lacy-burrows2020}.
Roman's coronagraph will act as a technology demonstrator, paving the way for a next generation of missions, expected to directly image low-mass exoplanets down to the Earth-like regime in reflected starlight.
With this goal, a Large IR/O/UV space telescope (recently named Habitable Worlds Observatory, HWO) was recommended to NASA as the next flagship mission by the US Astro 2020 Decadal Survey report \citep{astro2020}.
It combines some of the specifications of the LUVOIR\footnote{Large UV/Optical/IR Surveyor} \citep{bolcaretal2016, luvoirteam2018} and HabEx\footnote{Habitable Exoplanet Observatory} \citep{mennessonetal2016, gaudietal2018} concepts proposed to the US Decadal Survey.

The Large Interferometer For Exoplanets \citep[LIFE,][]{quanzetal2018, quanzetal2022} is a complementary approach to HWO, proposed to directly image the thermal emission of low-mass temperate exoplanets in the mid-infrared.
The mission concept consists of multiple free-flying telescopes operating together as a nulling interferometer in space.
The mid-IR range is suitable to characterize the atmospheres of Earth-like exoplanets due to the number of relevant molecules (e.g. H$_2$O, CO$_2$, O$_3$) with absorption bands at these wavelengths \citep{quanzetal2022b, konradetal2022, aleietal2022, angerhausenetal2022}.
This spectral range also provides a direct measurement of the planetary temperature and a strong constraint on its radius \citep{dannertetal2022}.
As such, mid-IR observations will overcome some of the physical degeneracies that affect the interpretation of reflected-starlight measurements, such as those caused by clouds and hazes \citep{lupuetal2016, fengetal2018} or by an unknown planet radius \citep{carriongonzalezetal2020}.
\citet{konradetal2022} discussed the prospects for LIFE to retrieve the atmospheric structure and composition of an Earth analogue with different configurations of signal-to-noise ratio (S/N), spectral resolution ($R$), wavelength coverage and mirror size of each of the four collecting telescopes of the interferometer.
They concluded that an accurate characterization could be achieved with a minimum wavelength coverage of 4$-$18.5 $\mu$m, $R$=50, S/N>10 and a mirror size of at least 2 m for a four-telescope configuration with a total 5\% instrument throughput.

A key to understanding the scientific potential of a given direct-imaging mission is to determine the parameter space of exoplanets to which the telescope is most sensitive.
Based on a synthetic population of exoplanets following occurrence rates from the \emph{Kepler} mission, \citet{kammerer-quanz2018} computed the possible exoplanet yield of a space-based mid-IR nulling interferometer and \citet{quanzetal2022} applied this methodology to the updated specifications of LIFE.
An interferometer with four telescopes of 2~m aperture and 5\% total throughput was found to yield more than 500 detectable exoplanets with planet radius 0.5$\,R_\oplus$<$R_p$<6$\,R_\oplus$ including between 10 and 20 exo-Earth candidates \citep[more details in Table 1 of][]{quanzetal2022}.
\citet{kammereretal2022} found that a configuration with four $\sim$3-m telescopes and 5\% throughput (or four $\sim$2-m telescopes and 20\% throughput) could detect about 30 rocky planets in the habitable zone of nearby Sun-like stars.

A different approach consists of using the current population of known exoplanets and determining the detectability of each planet with a given instrument, as done by \citet{brown2015} and \citet{carriongonzalezetal2021a} for reflected-starlight missions.
Although the set of known exoplanets remains biased due to the technology available --which is less sensitive to low-mass, long-period planets-- this approach provides a catalogue of already-existing targets.
This enables observational campaigns to improve the knowledge of their orbits and planetary systems, which will help in prioritizing targets.

Our goal in this work is to compute which of the currently known exoplanets are detectable with LIFE in the mid-IR.
We also compute their detectability with a notional performance estimate for the HWO direct-imaging mission observing in reflected starlight.
For that, we process the NASA Exoplanet Archive Database with the method described in \citet{carriongonzalezetal2021a}.
We analyse the overlap of LIFE's exoplanet yield with that of HWO in reflected starlight and with transit surveys.
This way we aim to provide a more concrete sense of the science cases that LIFE will be able to address, the targets that could benefit from multi-technique measurements, and those that are only detectable with LIFE.
We focus on the population of exoplanets within 20~pc from the Sun, which is expected to offer better opportunities for detection and atmospheric characterization \citep{quanzetal2022}.

This paper is structured as follows.
Section \ref{sec:methods} summarizes the methodology to compute orbital realizations for each of the known exoplanets and determine their detectability both with LIFE in the mid-IR and with HWO in reflected starlight.
Section \ref{sec:results} describes the set of exoplanets detectable with LIFE and compares it with those accessible with HWO. This section also identifies the exoplanets detectable only with LIFE, those that are also observable in transit, and those in the habitable zone (HZ) of their host stars.
The assumptions made throughout this work are discussed in Sect. \ref{sec:discussion}, and Sect. \ref{sec:conclusions} summarizes the main findings and conclusions.

\section{Methods} \label{sec:methods}
Below we describe the statistical methodology used to simulate the orbits of the currently confirmed exoplanets and derive some relevant orbital and planetary parameters (Sect. \ref{subsec:methods_orbits}).
Then, in Sects. \ref{subsec:methods_LIFE} and  \ref{subsec:methods_reflectedlight} we evaluate the detectability prospects of each planet with LIFE and with a reflected-starlight telescope.

\subsection{Simulated orbital realizations}  \label{subsec:methods_orbits}

Following the methodology in \citet{carriongonzalezetal2021a}, we downloaded the information of the known exoplanets from the NASA Exoplanet Archive\footnote{\url{https://exoplanetarchive.ipac.caltech.edu}} \citep{akesonetal2013}.
We used the Default Parameter Set in the NASA Exoplanet Archive's Planetary Systems database as of November 6, 2022, containing a total of 5197 confirmed exoplanets of which 259 are closer than 20~pc.
For some exoplanets we complemented this information with data from the original references therein, the Composite database of the NASA Exoplanet Archive or the Extrasolar Planets Encyclopaedia\footnote{\url{http://exoplanet.eu}} \citep{schneideretal2011}, as indicated in the text.

\begin{table}
\caption{Summary of the parameters used in our computations.}
\label{table:params_description}
\centering
    \begin{tabular}{c c l}
    \hline\hline
    Parameter   &   Unit    &   Description \\ 
    \hline
    $d$         &   pc      &   Distance to the star  \\
    $R_\star$   &   $R_\odot$   &   Stellar radius    \\
    $M_\star$   &   $M_\odot$   &   Stellar mass     \\
    $T_\star$   &   K       &   Stellar effective temperature   \\
    $a$         &   AU      &   Orbital semi-major axis of the planet  \\
    $P$         &   days    &   Orbital period      \\
    $T_{eq}$    &   K       &   Planet equilibrium temperature   \\
    $R_p$       &   $R_J$   &   Planet radius       \\
    $M_p$       &   $M_J$   &   Planet mass         \\
    $i$         &   deg     &   Orbital inclination     \\
    $\omega_p$    &   deg   &   Argument of periastron of the planet \\
    $e$         &   $-$     &   Orbital eccentricity  \\
\hline
\end{tabular}
\end{table}

For each exoplanet, we computed 1000 orbital realizations by randomly varying the orbital parameters (Table \ref{table:params_description}) within the uncertainties reported in the NASA Exoplanet Archive.
In those cases where the orbital information of a planet was incomplete, we used statistical arguments to have a complete description of the orbit.
For instance, we assumed isotropically distributed orbital orientations when the values of the orbital inclination ($i$) or the argument of periastron of the planet\footnote{\label{footnote_argper}We assumed that the reported values of the argument of periastron in the NASA Exoplanet Archive correspond to those of the star ($\omega_\star$) because that is the usual convention for radial velocity (RV) detections \citep[see e.g.][]{householder-weiss2022}. We refer to Sect. 4.1.1 in \citet{carriongonzalezetal2021a} for an additional discussion on the lack of a homogeneous convention in reporting $\omega$ and the relevance of this for accurately computing the detectability prospects.} ($\omega_p$) were not available.
In those cases we thus randomly drew values at each orbital realization from the uniform distributions $\cos(i)\in[-1,1]$ and $\omega_p\in[0,2\pi]$\, respectively.
With this we obtained a value of the planet mass ($M_p$) for each orbital realization even if only the minimum mass $M_p\,\sin(i)$ was available in the NASA Exoplanet Archive.
If the orbital eccentricity ($e$) was unknown we assumed a uniform probability distribution $e\in[0,1)$.
In case of an unknown planet radius ($R_p$) we computed it for each simulated orbit by means of previously published mass-radius relationships \citep{hatzes-rauer2015, otegietal2020}.
We refer to Sect. 4.2 in \citet{carriongonzalezetal2021a} for more information on the implementation of these relationships.

We discretize each orbital realization into 360 points evenly spaced in orbital true anomaly, and we compute the equilibrium temperature of the planet ($T_{eq}$) at each orbital position:
\begin{equation} \label{eq:Teq}
T_{eq} = \left( \frac{1-A_B}{4\,f} \right)^{1/4} \left( \frac{R_\star}{r} \right)^{1/2} T_\star.
\end{equation}
Here, $R_\star$ is the stellar radius, $T_\star$ is the stellar effective temperature, $r$ is the distance between the planet and the star at that orbital position and $A_B$ is the Bond albedo of the planet.
We fix $A_B$=0.45 for all planets to be consistent with our assumption of geometrical albedo $A_g$=0.3 and Lambertian scattering (see Sect. \ref{subsec:methods_reflectedlight}).
$f$ is a factor related to the heat redistribution of the planet for which we assume $f=1$, consistent with fast-rotating planets that have efficient heat redistribution \citep{traub-oppenheimer2010}.

In the cases where the values of $M_\star$, $R_\star$, or $T_\star$ were not available in the default parameter set of the NASA Exoplanet Archive, we obtained them from SWEET-Cat \citep{santosetal2013, sousaetal2021}.
When not available here either, we adopted the values reported in the NASA Archive from the Gaia DR2 \citep{gaia2018} or the Revised TESS Input Catalog \citep{stassunetal2019}.
In less than 1\% of cases, no values were to be found for these parameters and we excluded these planets from our analysis.

From this bootstrap-like method we derive statistical conclusions for the sampled parameters of each exoplanet.
In Table \ref{table:output_catalogue} we report the median values of the corresponding posterior distributions and their 16\% and 84\% confidence intervals (1$\sigma$ for Gaussian distributions), as resulting from our methodology, for all exoplanets within 20~pc.

\subsection{Detectability with LIFE} \label{subsec:methods_LIFE}
In this work we assume for LIFE an architecture consisting of a rectangular interferometric array with four free-flying telescopes sending their light to a fifth beam combiner spacecraft. Nulling of the central source (here the host star) is achieved with a dual chopped Bracewell beam combination scheme \citep{lay2004}. This architecture is particularly suitable for the suppression and self-calibration of instrumental noise as outlined in \citet{dannertetal2022}. We note that novel approaches such as three or five telescope kernel-nullers and their potential impact on the science return of a mission such as LIFE have been investigated in \citet{hansen-ireland2022, hansenetal2023}.

The integration time ($t_{int}$) required to detect a given planet is calculated using \texttt{LIFEsim} \citep{dannertetal2022}, which considers astrophysical noise sources such as stellar leakage, local zodiacal, and exozodiacal light, but does not account for instrumental noise sources as yet. This approach is only valid if LIFE is operated in a photon noise-dominated regime and Appendix~C of \citet{dannertetal2022} derives the minimum requirements on instrumental amplitude, phase, polarization, and spacecraft positioning errors to be operating in such a regime. We note that previous studies \citep{quanzetal2022,kammereretal2022} have assumed a conservative signal-to-noise ratio ($S/N$) of 7 (integrated over the full wavelength range) for a planet to be regarded as a detection to budget in additional instrumental noise. We adopted the same criterion in this work and assumed as detectable with LIFE any exoplanet for which an integrated $S/N$=7 is achieved in less than 100 hours of integration time.

For each known exoplanet, we provided LIFEsim with the following planetary and stellar parameters as inputs: planetary radius and equilibrium temperature, planet-star separation, distance to the planetary system and its ecliptic coordinates, and radius and effective temperature of the star.
The planetary parameters correspond to the median values obtained with our methodology (Table \ref{table:output_catalogue}) and the stellar parameters are obtained from the NASA Archive as indicated in Sect. \ref{subsec:methods_reflectedlight}.
In each case, we assumed the optimal instrument baseline that maximizes throughput at the angular separation of the planet semi-major axis. This is motivated by the fact that the considered planets are already known so that the observations can in principle be scheduled optimally.
In all cases, the nulling interferometry baseline length ranges between 10 and 100~m.

We note that the effective temperature of a planet may be higher than the $T_{eq}$ reported here due to, for instance, internal heat sources. These effects, are especially relevant for young exoplanets. Also, Solar System planets have shown that cold, mature giant planets can have internal heat sources resulting in effective temperatures significantly higher than the $T_{eq}$ \citep[e.g.][]{lietal2018}. This will increase the planet brightness in the mid-IR. A case-by-case analysis is needed, as the outcome is highly dependent on the evolutionary models and atmospheric properties assumed. Such analysis of the evolution of each particular planet is out of the scope of this work.

For our calculations, we consider three different scenarios to account for uncertainties in the instrumental parameters of LIFE. We consider mirror diameters of 1~m (pessimistic), 2~m (reference), and 3.5~m (optimistic) with their corresponding wavelength coverage of 6--17, 4--18.5, and 3--20~$\text{\textmu m}$ that will be adopted implicitly throughout this work.
Table~\ref{tab:life_mission_parameters} summarizes the mission parameters adopted for LIFE.
We refer to \citet{quanzetal2022} and \citet{dannertetal2022} for a more comprehensive discussion of their meaning and assumed values. Furthermore, we set the exozodiacal dust level to 3~zodi for all exoplanetary systems, motivated by the results from the HOSTS survey \citep{ertel2020}. In Sect. \ref{sec:discussion} we discuss the effects of assuming other values of exozodiacal dust.

For simplicity, we consider all stellar hosts as single. However, out of the 149 exoplanet-hosting stars within 20~pc, 30 of them --hosting a total of 49 exoplanets-- are multiples. Planets whose hosts are multiples might require longer integration times than what is shown here or might not be possible to be observed with LIFE at all (see details in Sect. \ref{sec:discussion}).
This will require detailed analyses for each case, to be done in future work.

\begin{table}
\caption{LIFE mission parameters adopted for the exoplanet yield predictions.}
\label{tab:life_mission_parameters}
\centering
\begin{tabular}{llcc}
\hline\hline
\multicolumn{2}{c}{Parameter} & Value & Unit\\
\hline
\multicolumn{2}{l}{Number of collector spacecraft} & 4 & --\\
Mirror diameter & pessimistic & 1.0 & m\\
\qquad \quad " & reference & 2.0 & m\\
\qquad \quad " & optimistic & 3.5 & m\\
Wavelength coverage & pessimistic & 6.0--17.0 & $\text{\textmu m}$\\
\qquad \quad " & reference & 4.0--18.5 & $\text{\textmu m}$\\
\qquad \quad " & optimistic & 3.0--20.0 & $\text{\textmu m}$\\
\multicolumn{2}{l}{Spectral resolution} & 20 & $\lambda/\Delta\lambda$\\ 
\multicolumn{2}{l}{Min. nulling baseline length} & 10 & m\\
\multicolumn{2}{l}{Max. nulling baseline length} & 100 & m\\
\multicolumn{2}{l}{Ratio imaging$/$nulling baseline} & 6 & --\\
\multicolumn{2}{l}{Quantum efficiency} & 0.7 & --\\
\multicolumn{2}{l}{Throughput} & 0.05 & --\\
\multicolumn{2}{l}{$\text{$S/N$}_\text{target}$} & 7 & --\\
\hline
\end{tabular}
\end{table}

\subsection{Detectability with reflected-starlight telescopes} \label{subsec:methods_reflectedlight}

Future direct imaging missions will achieve reflected-starlight observations of exoplanets in the visible and near-IR range by suppressing the stellar glare of the host star using devices such as coronagraphs or starshades.
Unavoidably, part of the inner region of the planetary system is obscured together with the host star, as defined by the inner working angle (IWA) of the instrument.
Similarly, these instruments have an outer working angle (OWA) that defines the outer limit of the detectable region. Beyond that, diffracted light from the suppression device reduces the contrast significantly.
In addition, the detectability is also limited by the minimum contrast of the instrument ($C_{min}$), such that the planet has to be brighter than $C_{min}$ in order to be distinguished from speckles and other noise.

The population of exoplanets accessible in reflected starlight is thus primarily defined by the IWA, the OWA, and $C_{min}$.
In this work, we considered these three parameters as the main detectability criteria for reflected-starlight observations.
Lacking an official list of specifications for the Habitable Worlds Observatory, we assumed notional values of IWA, OWA, and $C_{min}$ based on the Decadal Survey report and the LUVOIR and HabEx proposals.
Throughout this work, we assume a telescope diameter $D$=6~m, IWA=$3\, \lambda / D$, OWA=$64\, \lambda / D$, and $C_{min}$=$10^{-10}$.
We note, however, that a detailed instrument design and corresponding performance predictions for HWO are still work in progress.

For each orbital realization of a given exoplanet, we computed at each orbital position the angular separation ($\Delta \theta$) between the planet and the host star, and the planet brightness in reflected starlight given by the planet-to-star contrast ratio ($F_p / F_\star$).
We refer to Sect. 3 in \citet{carriongonzalezetal2021a} for more details on the computations.
In this work, we adopt a reference wavelength of $\lambda$=575~nm for all reflected starlight observations and a geometric albedo of $A_g$=0.3 for all exoplanets, with a Lambertian scattering phase law.

The orbital positions that are potentially observable in reflected starlight are hence those with $\Delta \theta$ between the IWA and the OWA, and $F_p/F_\star$ greater than $C_{min}$.
If these two criteria are simultaneously met, we refer to that exoplanet as accessible to the reflected-starlight mission under consideration.
The probability of the planet to be accessible to that mission ($P_{\rm{access}}$) is the ratio between the number of orbital realizations in which the planet is accessible and the total number of 1000 simulations.

Additional factors will also play a role in the detectability of exoplanets directly imaged in reflected starlight.
For instance, zodiacal and exozodiacal light might prevent the detection of certain targets.
The mission schedule and the pointing of the telescope will also restrict in practice the systems that can be observed at a given time \citep{brown2015}, although full sky coverage can be expected throughout a year for observatories located at the Earth-Sun L2 point.
The optical magnitude of the host star might affect the performance of the instrument and even prevent the observations of planets with faint host stars.
We also note that speckle subtraction techniques might improve the $C_{min}$ up to a factor of about 2.5 \citep[e.g.][]{ygoufetal2016}.
Evaluating the impact of these higher-order effects requires, however, a case-by-case analysis based on follow-up observations of each planet and instrument-specific tools to simulate the scheduling and the noise budget.
For this reason we generally use IWA, OWA, and $C_{min}$ as our main detectability criteria.

\section{Results} \label{sec:results}

\begin{figure*}
        \centering
        \includegraphics[width=8.cm]{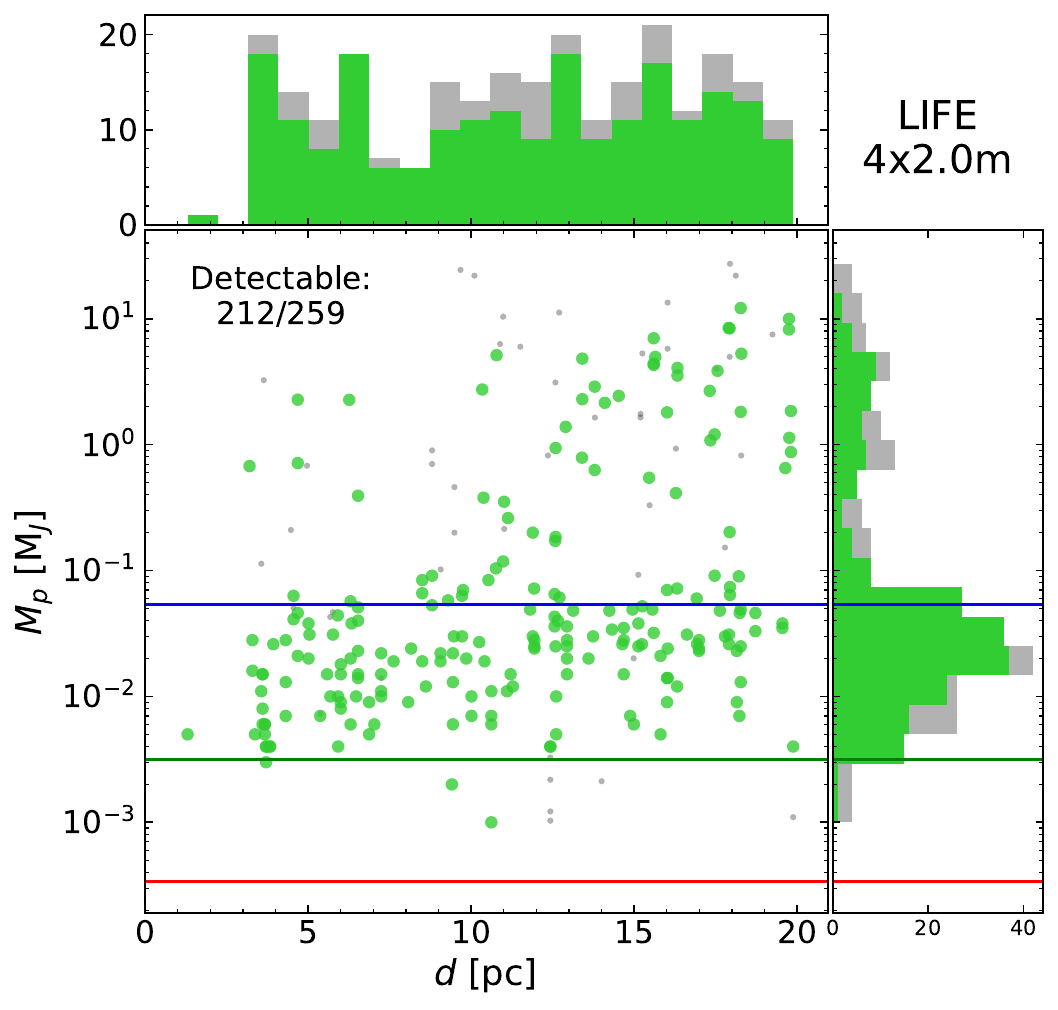} 
        \hfill
        \includegraphics[width=8.cm]{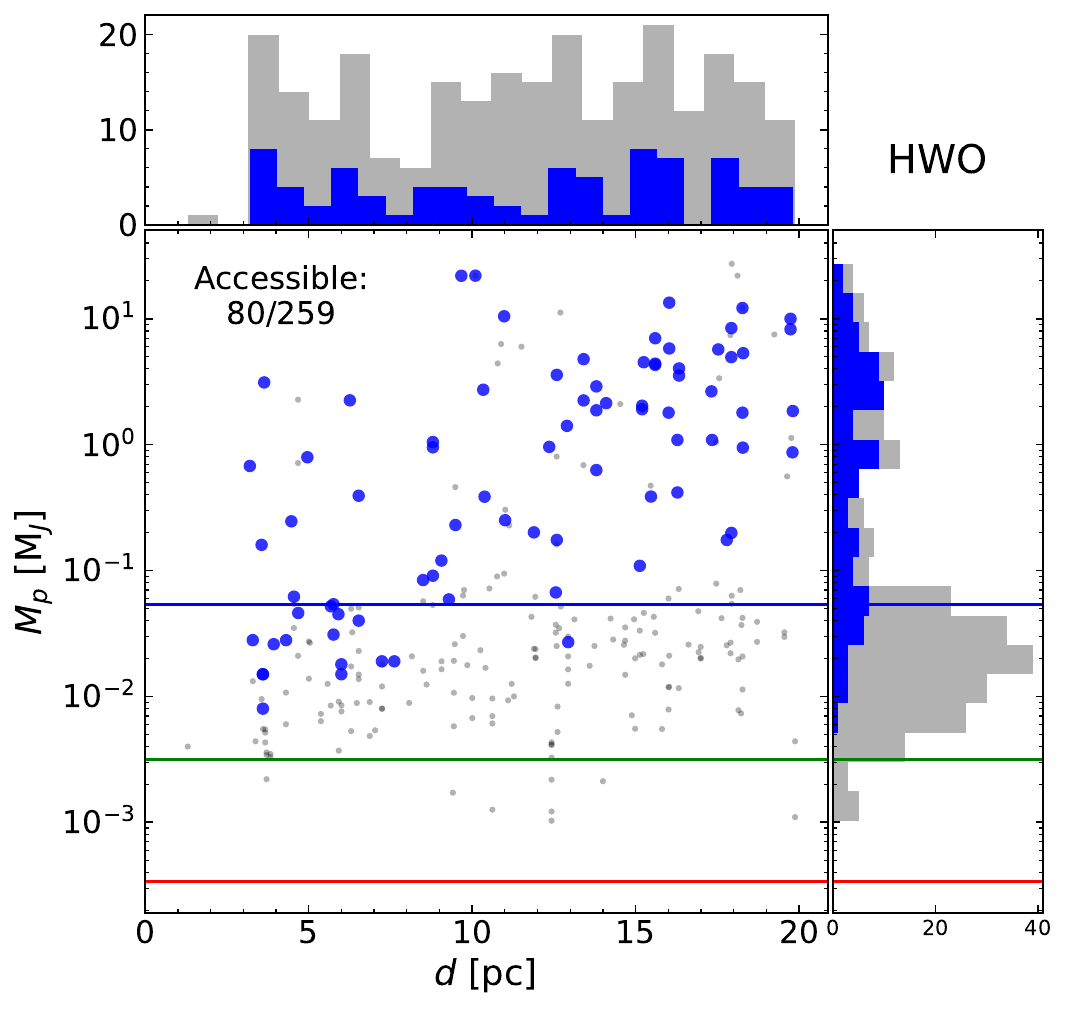} 
        \\
        \includegraphics[width=8.cm]{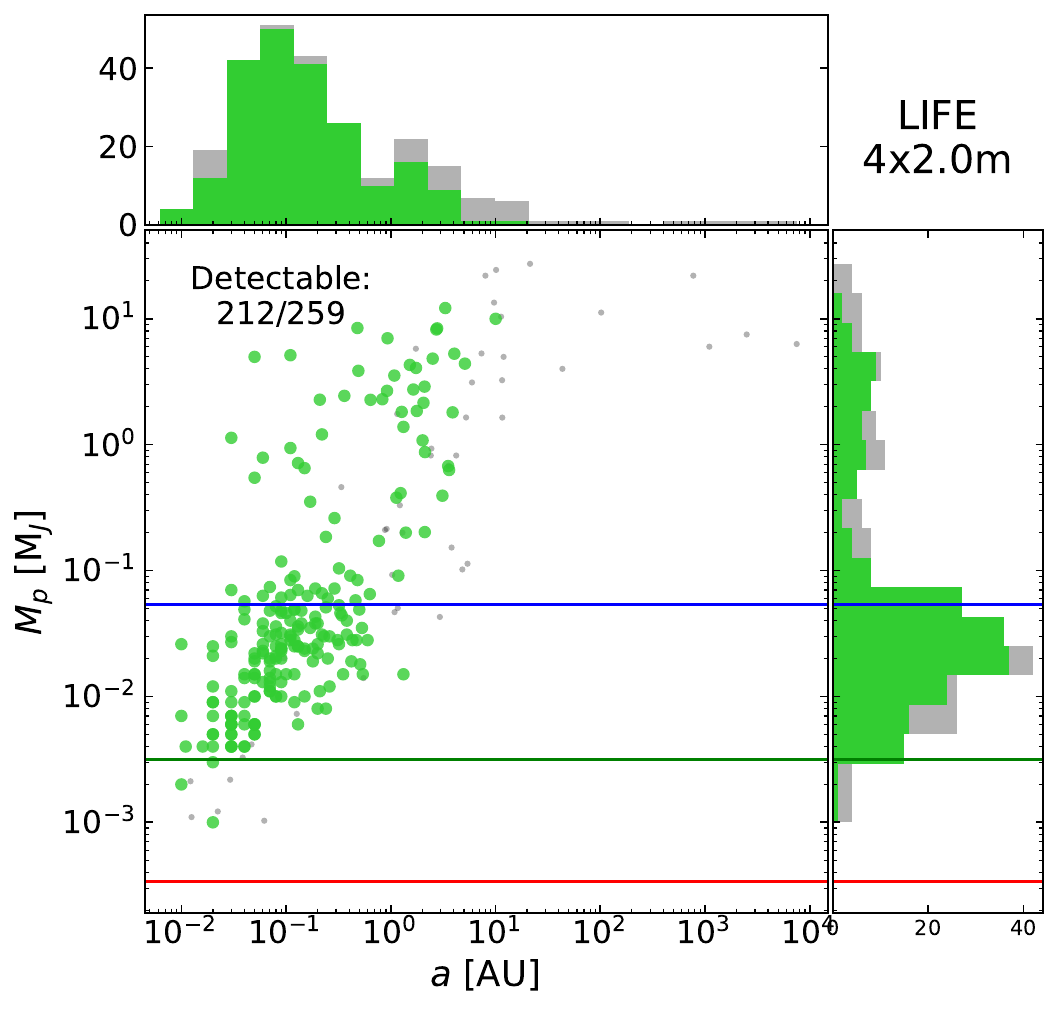} 
        \hfill
        \includegraphics[width=8.cm]{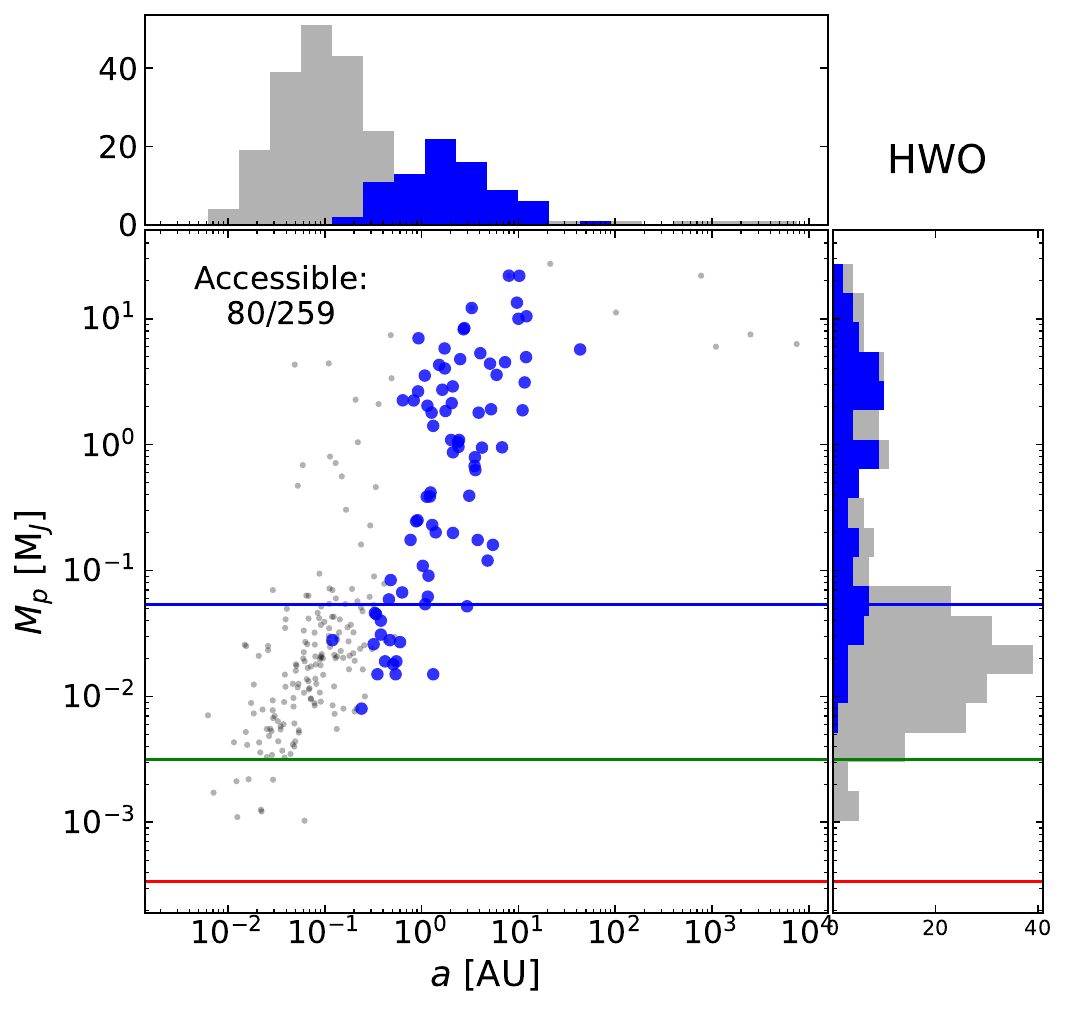} 

        \caption{\label{fig:detectableVSknown_all_missions} Potentially detectable currently-known exoplanets (coloured circles and histograms) compared to the total population of known exoplanets within 20~pc (grey dots and histograms), shown both as a function of their distance to the Solar System (top row) and their semimajor axis (bottom row). For LIFE (green), we assume the reference 4x2m configuration with 5\% throughput and consider planets as detectable if $t_{\rm{int}}(S/N=7)$<100~h. For the notional HWO (blue), potential targets in reflected starlight are those that meet our IWA, OWA and $C_{min}$ requirements (Sect. \ref{subsec:methods_reflectedlight}) and have $P_{\rm{access}}$>25\%. In all subplots, the value of $M_p$ reported for the potential targets is the output of our statistical methodology (Sect. \ref{subsec:methods_orbits}). For the non-detectable exoplanets we plot $M_p$ or alternatively $M_p\,sin\,i$ as reported in the NASA Archive. Horizontal lines indicate the masses of Neptune (blue), Earth (green) and Mars (red).}
\end{figure*} 
        
We show here the currently known planets within 20~pc that are detectable with LIFE, discussing the common targets for LIFE and HWO (Sect. \ref{subsec:results_reflected_light}) as well as those that are only detectable with LIFE and not accessible with HWO in reflected starlight (Sect. \ref{subsec:results_thermal_emission}).
Some of the planets detectable with LIFE have already been observed in transit and we comment on these in Sect. \ref{subsec:results_transiting}.
In Sect. \ref{subsec:results_habitable} we identify the exoplanets detectable with LIFE which orbit within the habitable zone of their host star.
Appendix \ref{sec:appendix_young} shows the integration times required to detect with LIFE the young exoplanets that have been directly imaged up to date.

\subsection{Synergies with reflected-starlight space missions} \label{subsec:results_reflected_light}

\begin{table*}
\tiny
\centering
\caption{Detectability of the exoplanets within 20 pc with the notional HWO mission in reflected starlight and LIFE in emitted light, computed with the methods in Sect. \ref{sec:methods}. For HWO, we include only planets with $P_{access}$>25\%. For LIFE, we list both the $t_{\rm{int}}$ required to achieve a detection (i.e. $S/N$=7) as well as the $S/N$ achieved in 10~h.
Integration times longer than 100~h are not shown.
}
\label{table:detectability_overlap_20pc} 
\begin{tabular}{l c c c c c c c c c c c c}
\hline 
\hline
& &  &  &  &  &	Notional HWO  & \multicolumn{2}{c}{LIFE (4x1m)}  & \multicolumn{2}{c}{LIFE (4x2m)}   &  \multicolumn{2}{c}{LIFE (4x3.5m)} \\ 
Planet & $d$ & 	 $a$ & 	 $R_p$ & $M_p$ & $T_{eq}$ & $P_{\rm{access}}$  & $S/N$ & $t_{\rm{int}}$  & $S/N$ & $t_{\rm{int}}$  & $S/N$ & $t_{\rm{int}}$  \\
& [pc] & [AU] & [$R_{\oplus}$] & [$M_{\oplus}$] & [K] &  [\%]  & (10~h)  &   [h] &   (10~h)  &   [h] &   (10~h)  &   [h] \\
\hline
tau Cet h & 	 3.60 & 	0.24 & 	1.35 & 	2.54 & 	417 & 	100.0 & 	41 & 	0.3  & 	107 & 	0.04  & 	219 & 	0.01 \\ 
tau Cet f & 	 3.60 & 	1.32 & 	1.91 & 	4.77 & 	175 & 	100.0 & 	 $-$ & 	 $-$  & 	7 & 	11  & 	18 & 	2 \\ 
tau Cet e & 	 3.60 & 	0.54 & 	1.91 & 	4.77 & 	278 & 	100.0 & 	25 & 	0.8  & 	77 & 	0.08  & 	174 & 	0.02 \\ 
HD 20794 d & 	 6.00 & 	0.35 & 	1.91 & 	4.77 & 	364 & 	100.0 & 	26 & 	0.7  & 	73 & 	0.09  & 	156 & 	0.02 \\ 
HD 20794 e & 	 6.00 & 	0.51 & 	2.24 & 	5.72 & 	311 & 	100.0 & 	23 & 	1.0  & 	69 & 	0.1  & 	153 & 	0.02 \\ 
GJ 514 b & 	 7.62 & 	0.42 & 	2.35 & 	6.04 & 	188 & 	99.8 & 	 $-$ & 	 $-$  & 	7 & 	10  & 	23 & 	0.9 \\ 
Kapteyn c & 	 3.93 & 	0.32 & 	2.80 & 	8.26 & 	142 & 	100.0 & 	 $-$ & 	 $-$  & 	8 & 	7  & 	29 & 	0.6 \\ 
HD 40307 g & 	 12.94 & 	0.60 & 	2.80 & 	8.90 & 	230 & 	32.2 & 	3 & 	68  & 	11 & 	4  & 	31 & 	0.5 \\ 
GJ 887 c & 	 3.29 & 	0.12 & 	2.91 & 	8.90 & 	331 & 	45.7 & 	155 & 0.02  & 	488 & 	0.002  & 	1132 & 	0.0004 \\ 
Wolf 1061 d & 	 4.31 & 	0.47 & 	2.91 & 	8.90 & 	134 & 	100.0 & 	 $-$ & 	 $-$  & 	4 & 	29  & 	15 & 	2 \\ 
GJ 229 A c & 	 5.76 & 	0.38 & 	3.14 & 	9.85 & 	191 & 	100.0 & 	6 & 	13  & 	26 & 	0.7  & 	82 & 	0.07 \\ 
HD 219134 g & 	 6.53 & 	0.38 & 	3.59 & 	12.71 & 	292 & 	100.0 & 	42 & 	0.3  & 	138 & 	0.03  & 	329 & 	0.005 \\ 
HD 180617 b & 	 5.91 & 	0.34 & 	3.92 & 	13.98 & 	174 & 	100.0 & 	5 & 	20  & 	22 & 	1  & 	69 & 	0.1 \\ 
GJ 876 e & 	 4.68 & 	0.33 & 	3.92 & 	14.62 & 	129 & 	100.0 & 	 $-$ & 	 $-$  & 	5 & 	21  & 	18 & 	2 \\ 
HD 102365 b & 	 9.29 & 	0.46 & 	4.60 & 	18.43 & 	355 & 	87.6 & 	69 & 	0.1  & 	214 & 	0.01  & 	485 & 	0.002 \\ 
HD 69830 d & 	 12.56 & 	0.63 & 	5.04 & 	20.66 & 	266 & 	100.0 & 	16 & 	2  & 	60 & 	0.1  & 	160 & 	0.02 \\ 
61 Vir d & 	 8.50 & 	0.48 & 	5.83 & 	26.70 & 	350 & 	100.0 & 	119 & 	0.03  & 	369 & 	0.004  & 	842 & 	0.0007 \\ 
HD 192310 c & 	 8.80 & 	1.18 & 	6.05 & 	28.92 & 	187 & 	100.0 & 	7 & 	9  & 	31 & 	0.5  & 	92 & 	0.06 \\ 
GJ 414 A c & 	 11.89 & 	1.39 & 	8.86 & 	63.56 & 	120 & 	100.0 & 	 $-$ & 	 $-$  & 	2 & 	83  & 	9 & 	6 \\ 
55 Cnc f & 	 12.58 & 	0.77 & 	9.19 & 	54.66 & 	238 & 	100.0 & 	30 & 	0.6  & 	115 & 	0.04  & 	319 & 	0.005 \\ 
HD 3765 b & 	 17.93 & 	2.11 & 	10.09 & 	64.20 & 	138 & 100.0 & 	 $-$ & 	 $-$  & 	4 & 	32  & 	14 & 	3 \\ 
pi Men b & 	 18.27 & 	3.31 & 	11.77 & 	3866.24 & 	193 & 100.0 & 	8 & 	9  & 	33 & 	0.5  & 	105 & 	0.04 \\ 
HD 160691 e & 	 15.60 & 	0.93 & 	12.11 & 	2224.72 & 	316 & 100.0 & 	126 & 	0.03  & 	421 & 	0.003  & 	1011 & 	0.0005 \\ 
14 Her b & 	 17.93 & 	2.77 & 	12.11 & 	2665.84 & 	139 & 100.0 & 	 $-$ & 	 $-$  & 	7 & 	11  & 	24 & 	0.9 \\ 
HD 87883 b & 	 18.29 & 	4.04 & 	12.33 & 	1677.75 & 	130 & 100.0 & 	 $-$ & 	 $-$  & 	3 & 	75  & 	10 & 	5 \\ 
ups And d & 	 13.41 & 	2.51 & 	12.33 & 	1533.15 & 	211 & 100.0 & 	26 & 	0.7  & 	97 & 	0.05  & 	255 & 	0.008 \\ 
HD 160691 b & 	 15.60 & 	1.52 & 	12.44 & 	1366.61 & 	245 & 100.0 & 	44 & 	0.3  & 	166 & 	0.02  & 	436 & 	0.003 \\ 
HD 128311 c & 	 16.33 & 	1.74 & 	12.44 & 	1293.20 & 	138 & 100.0 & 	 $-$ & 	 $-$  & 	7 & 	10  & 	24 & 	0.8 \\ 
HD 160691 c & 	 15.60 & 	5.10 & 	12.44 & 	1398.39 & 	134 & 100.0 & 	 $-$ & 	 $-$  & 	5 & 	24  & 	16 & 	2 \\ 
HD 128311 b & 	 16.33 & 	1.08 & 	12.55 & 	1124.43 & 	181 & 100.0 & 	8 & 	7  & 	36 & 	0.4  & 	118 & 	0.04 \\ 
47 UMa b & 	 13.79 & 	2.10 & 	12.67 & 	918.49 & 	184 & 100.0 & 	13 & 	3  & 	54 & 	0.2  & 	155 & 	0.02 \\ 
HR 810 b & 	 17.32 & 	0.92 & 	12.67 & 	850.48 & 	286 & 100.0 & 	81 & 	0.08  & 	282 & 	0.006  & 	702 & 	0.001 \\ 
HD 62509 b & 	 10.34 & 	1.64 & 	12.67 & 	871.45 & 	474 & 100.0 & 	144 & 	0.02  & 	334 & 	0.004  & 	654 & 	0.001 \\ 
gam Cep b & 	 14.10 & 	2.05 & 	12.78 & 	682.99 & 	314 & 100.0 & 	87 & 	0.06  & 	229 & 	0.009  & 	477 & 	0.002 \\ 
ups And c & 	 13.41 & 	0.83 & 	12.78 & 	729.39 & 	363 & 100.0 & 	255 & 	0.008  & 	742 & 	0.0009  & 	1600 & 	0.0002 \\ 
GJ 896 A b & 	 6.26 & 	0.64 & 	12.89 & 	721.76 & 	115 & 100.0 & 	3 & 	77  & 	13 & 	3  & 	51 & 	0.2 \\ 
HD 190360 b & 	 16.01 & 	3.89 & 	13.00 & 	573.66 & 	133 & 100.0 & 	 $-$ & 	 $-$  & 	5 & 	20  & 	19 & 	1 \\ 
HD 27442 b & 	 18.27 & 	1.27 & 	13.00 & 	578.74 & 	319 & 100.0 & 	84 & 	0.07  & 	246 & 	0.008  & 	539 & 	0.002 \\ 
7 CMa b & 	 19.81 & 	1.77 & 	13.00 & 	588.28 & 	333 & 100.0 & 	65 & 	0.1  & 	180 & 	0.02  & 	380 & 	0.003 \\ 
HD 147513 b & 	 12.90 & 	1.32 & 	13.11 & 	439.86 & 	218 & 100.0 & 	38 & 	0.3  & 	150 & 	0.02  & 	416 & 	0.003 \\ 
HD 10647 b & 	 17.34 & 	2.01 & 	13.23 & 	343.56 & 	193 & 100.0 & 	13 & 	3  & 	54 & 	0.2  & 	156 & 	0.02 \\ 
7 CMa c & 	 19.81 & 	2.12 & 	13.45 & 	277.45 & 	305 & 100.0 & 	55 & 	0.2  & 	158 & 	0.02  & 	344 & 	0.004 \\ 
bet Pic c & 	 19.75 & 	2.72 & 	13.45 & 	2610.86 & 	302 & 100.0 & 	73 & 	0.09  & 	214 & 	0.01  & 	466 & 	0.002 \\ 
47 UMa c & 	 13.79 & 	3.60 & 	13.67 & 	199.91 & 	141 & 100.0 & 	2 & 	93  & 	11 & 	4  & 	40 & 	0.3 \\ 
eps Eri b & 	 3.20 & 	3.53 & 	13.67 & 	214.84 & 	97 & 100.0 & 	 $-$ & 	 $-$  & 	9 & 	7  & 	30 & 	0.6 \\ 
HD 113538 b & 	 16.28 & 	1.24 & 	13.90 & 	130.94 & 	122 & 100.0 & 	 $-$ & 	 $-$  & 	4 & 	31  & 	15 & 	2 \\ 
HD 219134 h & 	 6.53 & 	3.11 & 	14.01 & 	124.90 & 	102 & 100.0 & 	 $-$ & 	 $-$  & 	5 & 	19  & 	18 & 	2 \\ 
GJ 649 b & 	 10.38 & 	1.13 & 	14.01 & 	120.45 & 	107 & 100.0 & 	 $-$ & 	 $-$  & 	4 & 	35  & 	14 & 	2 \\ 
bet Pic b & 	 19.75 & 	10.02 & 	18.49 & 	3175.94 & 	132 & 	100.0 & 	 $-$ & 	 $-$  & 	6 & 	12  & 	19 & 	1 \\ 

\hline
\end{tabular}
\end{table*}

\begin{figure*}
   \centering
   \includegraphics[width=16.cm]{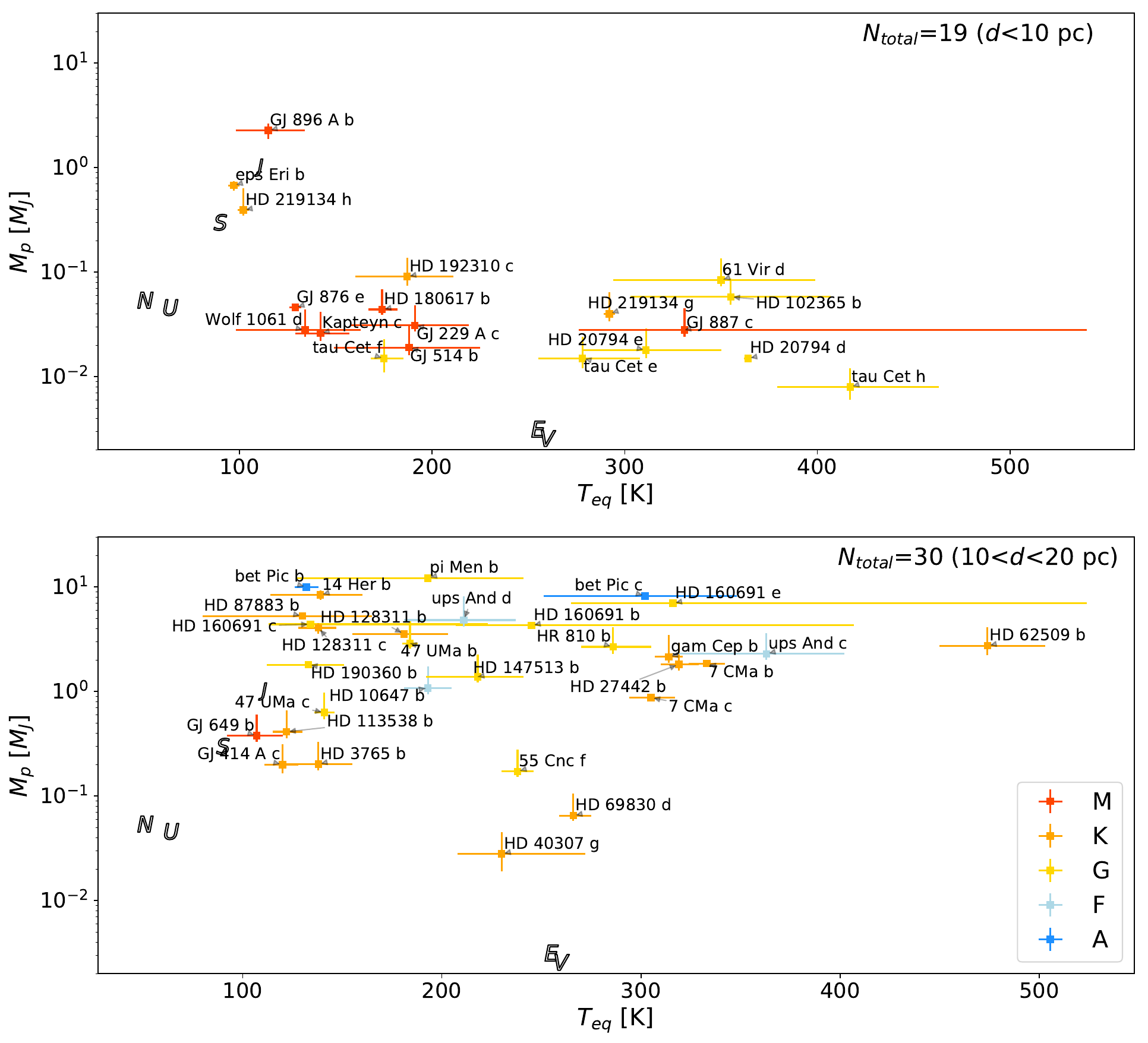}
      \caption{\label{fig:results_MpvsTeq_2m}
   Equilibrium temperatures and planet masses estimated for each of the currently-known exoplanets that are common targets for LIFE's reference configuration and the notional HWO.
   Each planet is colour-coded according to its host-star spectral type, as indicated in the legend.
   The top panel shows the common targets closer than 10 pc and the bottom panel shows those between 10 and 20 pc.
   $M_p$ and $T_{eq}$ are the median values computed from the 1000 orbital realizations (Sect. \ref{subsec:methods_reflectedlight}), with error bars corresponding to the upper and lower uncertainties (16\% and 84\% percentiles).
   Indicated with black letters are the Solar-System planets Venus (V), Earth (E), Jupiter (J), Saturn (S), Uranus (U), and Neptune (N).
      }
   \end{figure*}

\begin{figure}
   \centering
   \includegraphics[width=9.cm]{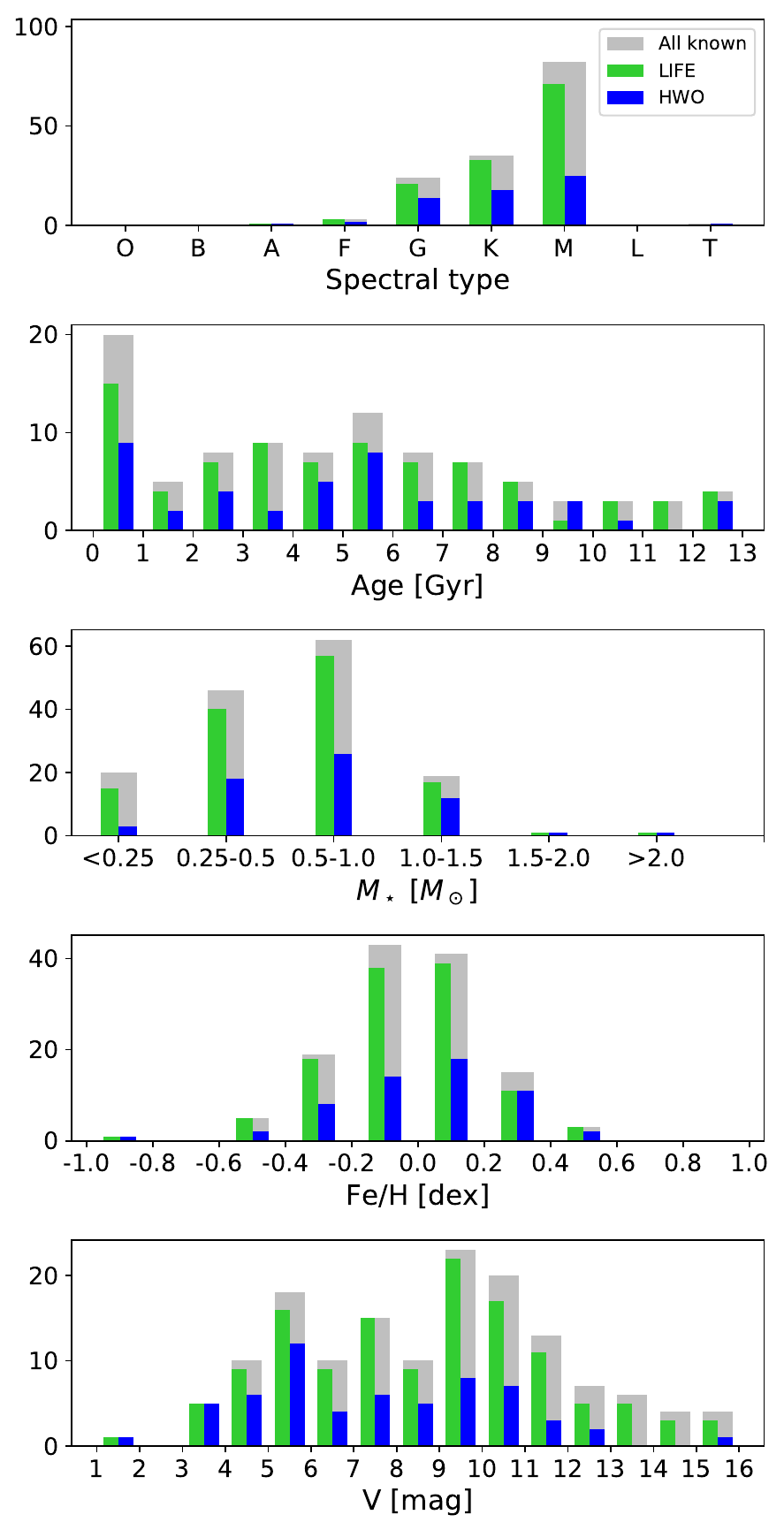}
      \caption{\label{fig:results_stellarproperties_2m}
      Stellar properties reported in the NASA Exoplanet Archive for the hosts of the LIFE-detectable exoplanets (4x2m and 5\% throughput) and those accessible in reflected-starlight with the notional HWO within 20~pc.
      Shown in grey for comparison, the stellar properties of the 149 stars hosting currently-known exoplanets within 20~pc.
      Out of these 149 stars, 97\% have available information on the spectral type, 63\% on the stellar age, 100\% on the stellar mass ($M_\star$), 85\% on the stellar metallicity, and 100\% on the optical magnitude ($V$).
      A total of 116 stars host LIFE-detectable exoplanets (resp. 61 for HWO). 
      }
   \end{figure}

Observing an exoplanet both at visible wavelengths and in the mid-IR will provide unique information about the planet and its atmosphere.
Visible, reflected-starlight observations are particularly sensitive to the optical properties of the clouds \citep[optical thickness, particle size and composition,][]{garciamunoz-isaak2015}, which provide key information on the meteorological phenomena taking place in the planet.
On the other hand, clouds trigger multiple degeneracies between physical parameters, hindering the interpretation of the measurements \citep{nayaketal2017, fengetal2018, damiano-hu2020, carriongonzalezetal2021b}.
The optical properties of the clouds are also highly degenerate with the value of the planet radius, if it is unknown \citep{carriongonzalezetal2020}.
If these parameter degeneracies are overcome, the visible wavelength range contains spectral features relevant to habitability studies, such as the ocean glint \citep{williams-gaidos2008, lustigyaegeretal2018} or the vegetation's red edge \citep{saganetal1993, seageretal2005}.
The mid-IR is less sensitive to the clouds and in principle offers less insight into their properties, but on the positive side it is less affected by the cloud-related degeneracies.
In general, the mid-IR can probe deeper down in the atmosphere and determine the temperature profile and the abundance of gaseous species \citep{quanzetal2022b, konradetal2022, aleietal2022, angerhausenetal2022}, several of which are potential biomarkers \citep{schwietermanetal2018}.

Figure \ref{fig:detectableVSknown_all_missions} shows the known exoplanets within 20~pc that are potential targets for LIFE and HWO, computed with the methodology described in Sect. \ref{sec:methods}.
Planets are considered detectable by LIFE if they have $t_{\rm{int}}$<100~h and they are considered accessible with HWO in reflected starlight if they meet $P_{\rm{access}}$>25\%.
A very high spatial resolution interferometric mission such as LIFE is particularly well suited to detect exoplanets orbiting at small separations.
This makes LIFE especially sensitive to the currently-known low-mass exoplanets, which have generally been found at small orbital separations due to current observational biases.
Indeed, LIFE can detect 144 of the 161 known planets (89\%) with $M_{p}$<$M_{Neptune}$ within 20~pc, and 68 of the 98 planets (69\%) with $M_{p}$>$M_{Neptune}$.
Reflected-starlight missions are less sensitive to the currently known low-mass planets.
The accessible planets within 20~pc for HWO are 16 with $M_{p}$<$M_{Neptune}$ (10\%) and 64 with $M_{p}$>$M_{Neptune}$ (65\%).
The reason for this is that most of the low-mass nearby planets discovered to date orbit M-type stars in short-period orbits which fall within the IWA of optical coronagraphs.
This gap in the parameter space of currently-discovered exoplanets is clearly shown in the bottom row of Fig. \ref{fig:detectableVSknown_all_missions}.

Figure \ref{fig:detectableVSknown_all_missions_beyond20pc} in Appendix \ref{sec:appendix_beyond20pc} extends Fig. \ref{fig:detectableVSknown_all_missions} to show the planets beyond 20~pc accessible with HWO in reflected starlight.
Our criteria to deem a planet accessible in reflected starlight (IWA, OWA, $C_{min}$) are mostly geometrical and thus less restrictive than those to deem a planet detectable with LIFE, because \texttt{LIFEsim} includes additional noise sources and instrument sensitivity.
Based on these geometrical considerations, we find a large number of long-period giant planets accessible to HWO out to about 100~pc (Fig. \ref{fig:detectableVSknown_all_missions_beyond20pc}).
Whether these targets will be actually detectable will depend on the final operational and performance specifications of the mission and its sensitivity limits, which are not yet defined.
Nevertheless, the number of low-mass planets accessible in reflected starlight beyond 20~pc does not increase from those in Fig. \ref{fig:detectableVSknown_all_missions}.

Table~\ref{table:detectability_overlap_20pc} shows the list of 55 exoplanets within 20~pc that are detectable with LIFE's reference configuration and are also accessible with the notional HWO mission in reflected starlight.
Figure~\ref{fig:results_MpvsTeq_2m} provides additional information on the equilibrium temperature and mass of these planets.
We find that most of the planets that are common targets for LIFE and HWO are cold and temperate giant planets, with masses similar to that of Jupiter and equilibrium temperatures between that of Jupiter and that of the Earth.
Within 10~pc, some planets less massive than Neptune are accessible for both LIFE and HWO, including the low-mass planets tau Cet e, f and h, HD 20794 d and e, and GJ 514 b.

Figure \ref{fig:results_stellarproperties_2m} compares the properties of the stars within 20~pc hosting potential targets for the notional LIFE and HWO missions.
Among the known exoplanet-hosts, LIFE can study planetary systems with hosts of practically all stellar types and masses, and especially with stars less massive than the Sun.
The information on stellar age and metallicity is scarcer in the NASA Archive (e.g. only available for 59\% and 87\% of LIFE-detectable-planet hosts, respectively).
However, we find that LIFE may analyse a significant number of planets orbiting stars of almost any age or metacillity value.
We note that M-type stars frequently lack an exact value of their age in the NASA Archive, which especially affects the tally of LIFE-detectable-planet hosts.
LIFE's mid-IR wavelength range also makes it more sensitive to stars that are dim in the visible ($V\gtrsim9$ mag).
The $V$ magnitude of a star will be a limiting factor for the detectability of exoplanets with reflected-starlight missions, as indicated in Sect. \ref{subsec:methods_reflectedlight}.

The exoplanets discussed in this section are valuable candidates for an in-depth atmospheric characterization spanning a broad range of wavelengths.
In addition, large numbers of planets are expected to be discovered in the coming years and decades with ongoing and upcoming transit and radial-velocity instruments.
As shown in Fig. \ref{fig:detectableVSknown_all_missions}, LIFE performs well over a broad parameter space and will be able to detect a significant fraction of these newly discovered targets.
The particularly interesting cases of temperate long-period planets around nearby Sun-like stars will generally be accessible both with LIFE and with HWO in reflected starlight.
The number of overlapping targets is thus expected to increase in the near future.

\subsection{Unique parameter space for LIFE} \label{subsec:results_thermal_emission}

\begin{figure}
   \centering
   \includegraphics[width=9.cm]{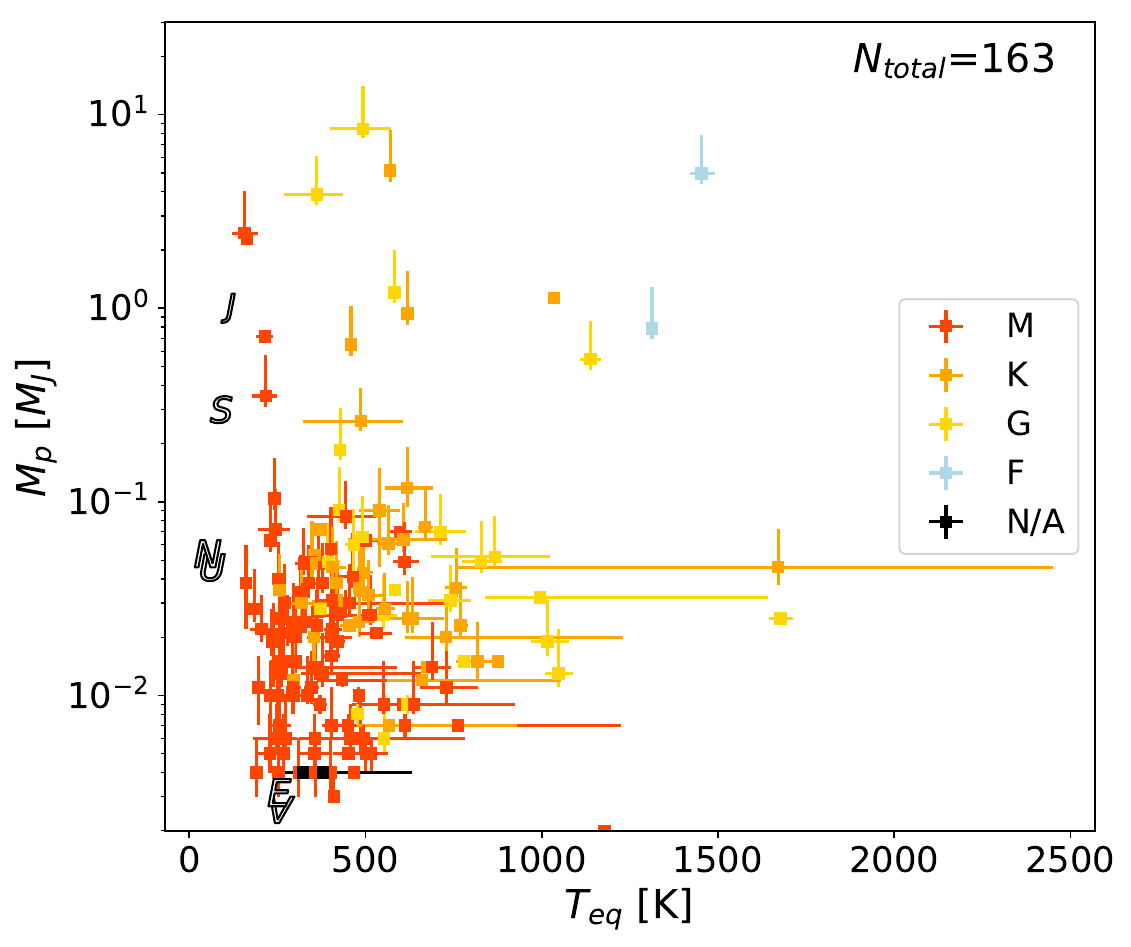}
   \caption{\label{fig:results_MpvsTeq_2m_onlyLIFE}
   Equilibrium temperatures and planet masses estimated for each of the known exoplanets within 20~pc that are detectable with LIFE (4x2m and 5\% throughput) but not with the notional HWO. Colour-coding corresponds to the host-star spectral type. Table \ref{table:detectability_onlyLIFE_20pc} contains the list of planets and their properties.
      }
   \end{figure}

The nulling interferometry method allows LIFE to detect planets on short-period orbits that cannot be resolved with optical coronagraphs or starshades.
Also, with temperate exoplanets' thermal emission peaking in the mid-IR, LIFE is more sensitive to low-mass planets that might not reflect enough light to be detectable in the visible.
Figure \ref{fig:results_MpvsTeq_2m_onlyLIFE} shows the 163 known exoplanets within 20~pc that can only be detected with LIFE's reference configuration and not with HWO in reflected starlight.
Table~\ref{table:detectability_onlyLIFE_20pc} displays additional details on the main properties of these planets.
Appendix Fig. \ref{fig:results_MpvsTeq_2m_onlyHWO} shows the 31 known exoplanets within 20~pc accessible in reflected starlight to the notional HWO but not detectable with LIFE's reference configuration --giant planets with low $T_{eq}$ which do not emit enough thermal radiation to allow a detection in the mid-IR.

In the reference LIFE configuration (4x2~m telescopes with 5\% throughput), only eight of these 163 planets require integration times longer than 10 hours to be detected with a $S/N$ of 7. 
Among the targets only detectable by LIFE, we find 62 exoplanets with masses $M_p$<5$M_\oplus$. 
Of these low-mass planets, 37 require one hour of integration or less in order to achieve a detection with $S/N$=7 in the reference configuration.
Appendix Fig. \ref{fig:detectable_vs_known_LIFE_SNR-coloured} shows the $t_{int}$ of each LIFE-detectable exoplanet.

The variety of exoplanets only detectable with LIFE ranges from hot Jupiters (e.g. 51 Peg b) to temperate planets with similar mass to that of the Earth (e.g. Teegarden's Star b and c).
The host stars of these exoplanets also range from stars more massive than the Sun down to M dwarfs.
Below we comment in more detail on the planets detectable with LIFE which have also been detected in transit (Sect. \ref{subsec:results_transiting}) and those that orbit within the habitable zone of their host star (Sect. \ref{subsec:results_habitable}).

\subsection{Transiting planets detectable with LIFE} \label{subsec:results_transiting}

\begin{table}
\tiny
\centering
\caption{Integration times required to detect with $S/N$=7 the known transiting exoplanets within 20~pc, for each of the LIFE configurations assuming a total 5\% throughput in all cases.}
\label{table:results_transiting}
\begin{tabular}{l c c c c c c}
    \hline
    \hline
    
    Planet & $d$ & 	 $a$ & 	 $R_p$    &   4x1 m   &   4x2 m   &   4x3.5 m \\
            & [pc] & [AU] & [$R_{\oplus}$] &   [h]     &   [h]     &   [h]     \\
    \hline
HD 189733 b & 	 19.76 & 	0.03 & 	12.67 & 	0.004  &  0.0003  &  6e-05  \\ 
GJ 436 b & 	 9.75 & 	0.03 & 	4.15 & 	0.02  &  0.003  &  0.0008  \\ 
AU Mic b & 	 9.72 & 	0.06 & 	4.71 & 	0.02  &  0.004  &  0.001  \\ 
HD 219134 b & 	 6.53 & 	0.04 & 	1.57 & 	0.1  &  0.02  &  0.006  \\ 
HD 219134 c & 	 6.53 & 	0.07 & 	1.46 & 	0.2  &  0.03  &  0.01  \\ 
HD 136352 c & 	 14.68 & 	0.17 & 	2.91 & 	0.2  &  0.03  &  0.007  \\ 
AU Mic c & 	 9.72 & 	0.11 & 	2.91 & 	0.2  &  0.02  &  0.004  \\ 
HD 136352 b & 	 14.68 & 	0.10 & 	1.68 & 	0.7  &  0.1  &  0.03  \\ 
GJ 357 b & 	 9.44 & 	0.04 & 	1.23 & 	0.9  &  0.1  &  0.02  \\ 
pi Men c & 	 18.27 & 	0.07 & 	1.79 & 	1  &  0.1  &  0.04  \\ 
55 Cnc e & 	 12.59 & 	0.02 & 	1.91 & 	1  &  0.09  &  0.02  \\ 
LTT 1445 A c & 	 6.87 & 	0.03 & 	1.12 & 	1  &  0.2  &  0.03  \\ 
HD 260655 c & 	 10.01 & 	0.05 & 	1.57 & 	1  &  0.2  &  0.04  \\ 
GJ 486 b & 	 8.07 & 	0.02 & 	1.35 & 	1  &  0.2  &  0.04  \\ 
LTT 1445 A b & 	 6.87 & 	0.04 & 	1.35 & 	2  &  0.2  &  0.05  \\ 
GJ 143 b & 	 16.32 & 	0.19 & 	2.58 & 	3  &  0.3  &  0.05  \\ 
HD 260655 b & 	 10.01 & 	0.03 & 	1.23 & 	3  &  0.4  &  0.1  \\ 
HD 136352 d & 	 14.68 & 	0.43 & 	2.58 & 	4  &  0.4  &  0.07  \\ 
HD 21749 c & 	 16.32 & 	0.07 & 	0.90 & 	6  &  0.8  &  0.2  \\ 
L 98-59 d & 	 10.62 & 	0.05 & 	1.57 & 	11  &  1  &  0.2  \\ 
L 98-59 c & 	 10.62 & 	0.03 & 	1.35 & 	14  &  1  &  0.3  \\ 
LHS 3844 b & 	 14.88 & 	0.01 & 	1.35 & 	26  &  0.3  &  0.05  \\ 
GJ 1132 b & 	 12.61 & 	0.02 & 	1.12 & 	42  &  2  &  0.4  \\ 
L 98-59 b & 	 10.62 & 	0.02 & 	0.90 & 	51  &  4  &  1  \\ 
GJ 367 b & 	 9.41 & 	0.01 & 	0.67 & 	63  &  5  &  1  \\ 
LHS 1478 b & 	 18.23 & 	0.02 & 	1.23 & 	105  &  2  &  0.5  \\ 
GJ 1214 b & 	 14.64 & 	0.01 & 	2.69 & 	157  &  4  &  0.8  \\ 
GJ 3929 b & 	 15.81 & 	0.03 & 	1.12 & 	165  &  9  &  2  \\ 
LHS 1678 c & 	 19.88 & 	0.03 & 	1.01 & 	1116  &  50  &  10  \\ 
LHS 1140 c & 	 14.99 & 	0.03 & 	1.12 & 	1158  &  51  &  8  \\ 
TRAPPIST-1 c & 	 12.43 & 	0.016 & 	1.12 & 	2202  &  73  &  10  \\ 
TRAPPIST-1 b & 	 12.43 & 	0.011 & 	1.12 & 	4558  &  97  &  15  \\ 
TOI-540 b & 	 14.00 & 	0.01 & 	0.90 & 	7493  &  146  &  32  \\ 
LHS 1678 b & 	 19.88 & 	0.01 & 	0.67 & 	13588  &  238  &  52  \\ 
LHS 1140 b & 	 14.99 & 	0.10 & 	1.68 & 	44160  &  2596  &  268  \\ 
TRAPPIST-1 e & 	 12.43 & 	0.029 & 	0.90 & 	83523  &  4681  &  526  \\ 
TRAPPIST-1 d & 	 12.43 & 	0.022 & 	0.78 & 	100405  &  4712  &  575  \\ 
TRAPPIST-1 f & 	 12.43 & 	0.038 & 	1.01 & 	115282  &  6872  &  736  \\ 
TRAPPIST-1 g & 	 12.43 & 	0.047 & 	1.12 & 	132300  &  7835  &  813  \\ 
TRAPPIST-1 h & 	 12.43 & 	0.062 & 	0.78 & 	--  &  166656  &  16422  \\ 
\end{tabular}
\end{table}

As of November 6, 2022, 40 of the known exoplanets within 20~pc have been found to transit their host stars.
These planets are listed in Table \ref{table:results_transiting} together with the integration times required to detect them with different configurations of LIFE.

LIFE will be able to detect 32 of these transiting planets with $t_{int}$<100~h in the reference 4x2~m configuration with 5\% throughput.
This is a result of the high angular resolution achievable with a space-based mid-IR interferometer. 
None of these 40 planets is accessible with HWO in reflected starlight due to their small orbital separations.

The LIFE-detectable transiting planets in Table \ref{table:results_transiting} are mainly warm and hot ($T_{eq} \sim 500-700$ K) short-period planets in the super-Earth to Neptune-like mass regime.
Several hotter planets are also detectable, such as the hot-Jupiter HD 189733 b ($T_{eq}$=1036~K) and the super-Earths 55 Cnc e ($T_{eq}$=1678~K) and pi Men c ($T_{eq}$=1006~K).
These planets orbit too close to their host stars to be resolved by optical coronagraphs.

Multi-planet systems are valuable case studies to understand planetary systems as a whole and to constrain the possible formation and evolution of the planets.
We find several systems with more than one transiting planet detectable with LIFE.
These are HD 219134 (b, c), L 98-59 (b, c, d), HD 136352 (b, c, d), GJ 143 --also named HD 21749-- (b, c), HD 260655 (b, c), LTT 1445 A (b, c), AU Mic (b, c), and TRAPPIST-1 (b, c).
Several other multi-planet systems with one or no transiting planets are also detectable with LIFE, as shown in Tables \ref{table:detectability_overlap_20pc} and \ref{table:detectability_onlyLIFE_20pc}.

Among these known transiting planets, the only ones in the habitable zone of their host star are LHS 1140 b and TRAPPIST-1 e, f, g.
Although these planets are interesting for habitability studies \citep[see e.g.][]{wunderlichetal2021}, their $t_{int}$ are beyond our detectability criteria even for LIFE's optimistic configuration.
This is due to the extremely small angular separation of these planets around stars of very low mass (e.g. 1.6 and 2.4 milliarcseconds for TRAPPIST-1 e and f, respectively).
The nulling baseline length needed to optimally resolve these planets is larger than the maximum baseline length of 100~m assumed in this work (Table \ref{tab:life_mission_parameters})\footnote{We recall that the first positive transmission peak in the nulling interferometer is proportional to the nulling baseline $b$, $\lambda / 2b$, while for a single-mirror telescope the spatial resolution is proportional to the mirror diameter $D$, $\lambda / D$ \citep[see e.g.][]{dannertetal2022}.}.
The observation of these planets cannot be optimized, resulting in a significant stellar leakage that prevents a detection in $t_{int}$<100~h.
Indeed, we verified that $t_{int}$ drops to about 100~h for these planets with a nulling baseline of 500~m.

The list of transiting exoplanets will grow in the next years with missions monitoring nearby systems such as TESS and PLATO.
These missions will discover long-period transiting planets accessible to LIFE and also to reflected-starlight imaging telescopes, constraining the value of their radius.
Nearby transiting exoplanets will also be targeted by JWST and Ariel, and many of those in Table \ref{table:results_transiting} will be thoroughly studied in the coming years.
Several of them have been observed with ground-based high-resolution Doppler spectroscopy \citep[e.g.][]{bourrieretal2022, lopezmoralesetal2014, wyttenbachetal2015}.
It thus unlikely that planets in Table \ref{table:results_transiting} will be prime targets for LIFE, although they illustrate the capabilities of the mission.
We will study in future work the possible improvement in the atmospheric characterization if transit and direct-imaging observations are combined.

\subsection{Habitable zone planets}  \label{subsec:results_habitable}

\begin{table}
\centering
\small
\caption{Detectability of known habitable-zone planets within 20~pc with LIFE and the HWO reflected-starlight notional mission. For LIFE, the total sum of integration time needed to detect those planets ($\sum t_{int}$ [h]) is also indicated.}
\label{table:results_HZ_count} 
\resizebox{!}{1.3cm}{
\begin{tabular}{l c c c c c c c }
\hline
\hline
    &   \multicolumn{2}{c}{LIFE 4x1~m}    &   \multicolumn{2}{c}{LIFE 4x2~m}    &   \multicolumn{2}{c}{LIFE 4x3.5~m}  & HWO    \\
    &   Tally &  $\sum t_{int}$ &      Tally &  $\sum t_{int}$   &      Tally &  $\sum t_{int}$ &          \\
\hline
Total HZ                    &   32 & 424  &   38 & 138   &   38 & 15     &       13     \\
<5$M_\oplus$                &   9 & 94    &   13 & 77    &   13 & 8      &        1     \\
5$M_\oplus$-10$M_\oplus$   &    6 & 212  &    8 & 54    &   8 & 6       &        2    \\
10$M_\oplus$-1$M_{J}$    &   11 & 111  &   11 & 7     &   11 & 0.8    &        4  \\
>1$M_{J}$                 &    6 & 7    &    6 & 0.4   &   6 & 0.06    &        6   \\
\hline
\end{tabular}
}
\end{table}

\begin{figure*}
   \centering
   \includegraphics[width=8.1cm]{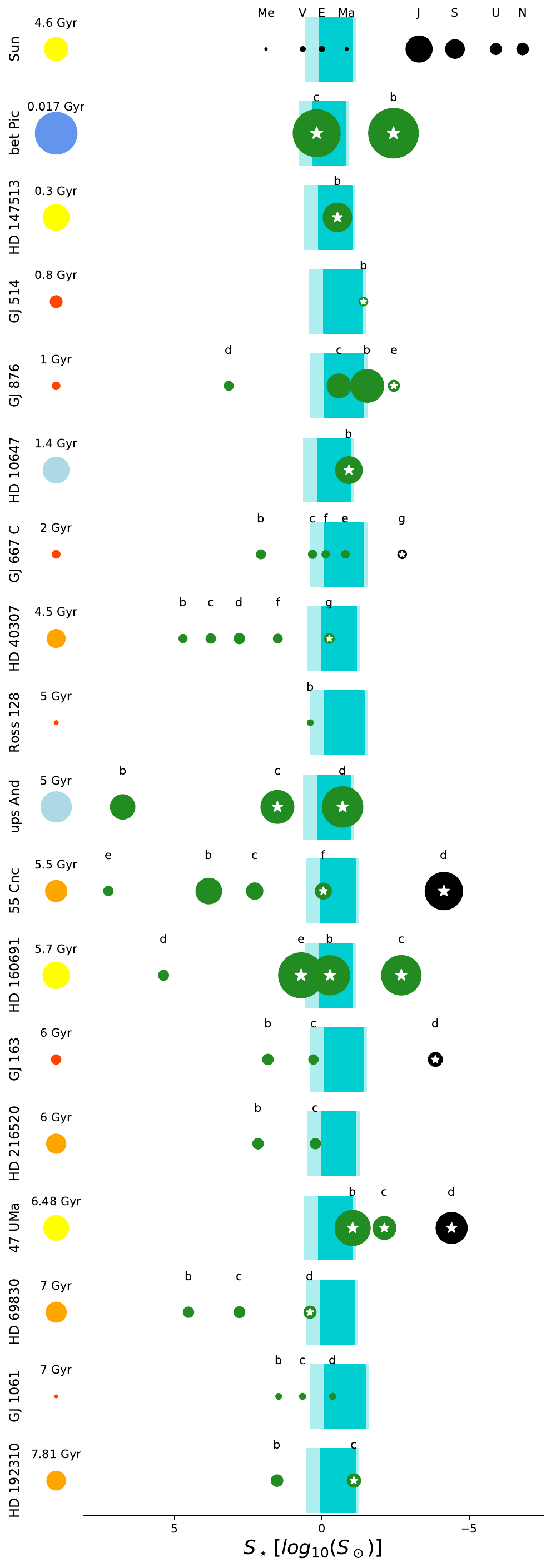}
   \hfill
   \includegraphics[width=8.1cm]{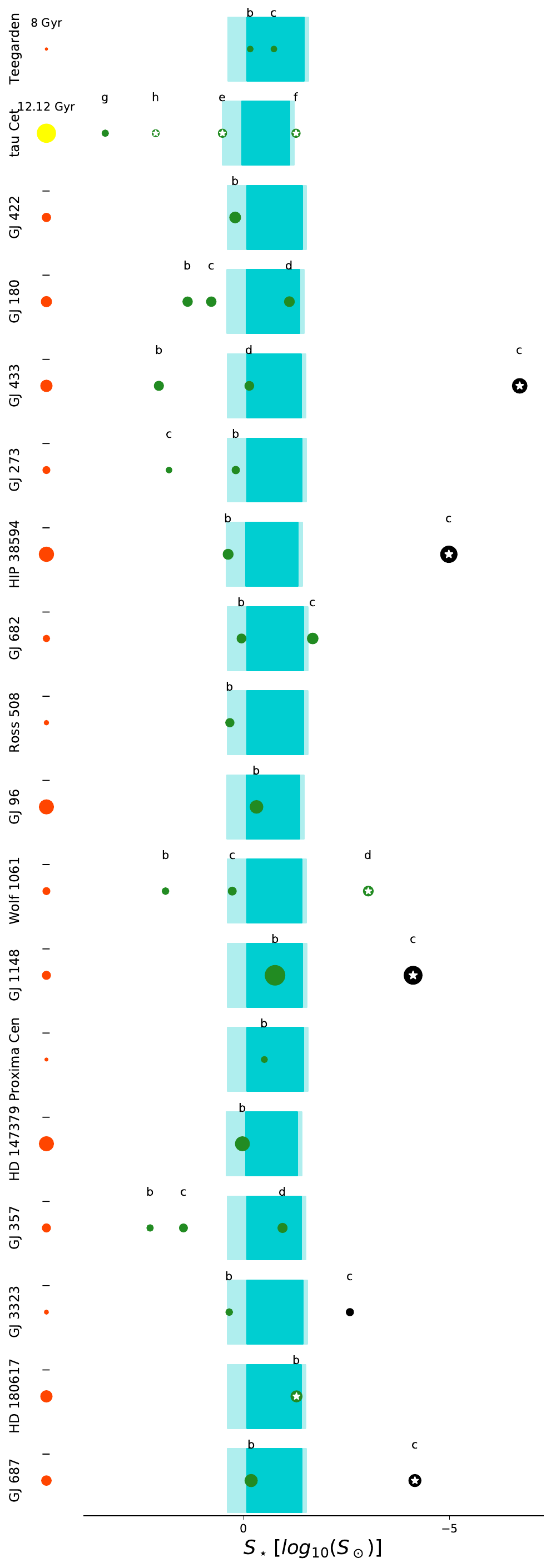}
      \caption{\label{fig:results_HZ1_2m} Planetary systems with HZ planets detectable with LIFE's reference configuration. Planets detectable with LIFE are shown by green circles and those which are not, by black circles. Inscribed white-star markers indicate planets that are accessible with HWO in reflected starlight ($P_{\rm{access}}$>25\%). Blue-shaded regions show the conservative HZ of each star, with lighter-blue regions being the optimistic HZ zone \citep{kopparapu2014}. The sizes of the markers both for stars and planets are proportional to their masses. Stars are color-coded depending on their spectral type (red, orange, yellow, blue-grey and light blue for M, K, G, F and A stars resp.). The Solar System is depicted in the first subplot for reference.} 
\end{figure*}

One of the main goals of the next generation of space telescopes is to characterize the atmospheres of exoplanets in the habitable zone of their host stars. 
Although the number of such planets discovered to date is still scarce due to the sensitivity biases of current detection methods, several dozen have been already confirmed.

Here we discuss the known HZ planets within 20~pc detectable with LIFE, as well as those that are also accessible in reflected starlight to HWO.
Figure \ref{fig:results_HZ1_2m} shows all the systems with HZ-planets detectable ($t_{int}$<100~h) with LIFE's reference configuration (green circles). 
We also show in that figure the planets that are accessible to HWO ($P_{\rm{access}}$>25\%), as a representative configuration of future reflected-starlight missions.
Table \ref{table:results_HZ_count} provides the tally of detectable planets for the three configurations of LIFE and HWO.
It also shows the total sum of integration time ($\sum t_{int}$) required for detecting (broadband $S/N$=7) the LIFE-detectable HZ planets.
We consider both the optimistic and conservative HZ as defined by \citet{kopparapu2014}.
In all cases, we take as reference the HZ for a 1$M_\oplus$ planet computed with their Eq. (4).

A total of 38 known HZ planets are detectable with LIFE's reference configuration.
For 13 of these planets, we estimate masses lower than 5$M_\oplus$ (tau Cet e, GJ 667 C c, e and f, Teegarden's Star b and c, Ross 128 b, GJ 273 b, Ross 508 b, Wolf 1061 c, Proxima Centauri b, GJ 3323 b, and GJ 1061 d).
We find 8 planets with masses between 5$M_\oplus$--10$M_\oplus$ (GJ 180 d, GJ 433 d, HIP 38594 b, GJ 682 b, GJ 163 c, GJ 514 b, GJ 357 d, and HD 40307 g).
We also find 11 planets with masses between 10$M_\oplus$--1$M_{Jup}$ (HD 69830 d, GJ 422 b, HD 192310 c, GJ 96 b, GJ 1148 b, HD 180617 b, HD 147379 b, 55 Cnc f, GJ 876 c, GJ 687 b, and HD 216520 c).
Finally, 6 planets are found with $M_p$>1$M_{Jup}$ (HD 160691 b, 47 UMa b, bet Pic c, HD 10647 b, HD 147513 b, and ups And d).
In summary, out of the total 38 LIFE-detectable HZ planets, 34\% of them have $M_p$<5$M_\oplus$, 21\% have 5$M_\oplus$<$M_p$<10$M_\oplus$, 29\% have 10$M_\oplus$<$M_p$<1$M_{Jup}$ and 16\% have $M_p$>1$M_{Jup}$.

For comparison, 13 of the HZ-planets within 20~pc are accessible to HWO in reflected starlight.
Of these, one has a mass lower than 5$M_\oplus$ (tau Cet e), two have mass between 5$M_\oplus$--10$M_\oplus$ (HD 40307 g, GJ 514 b), four have masses between 10$M_\oplus$--1$M_{Jup}$ (HD 69830 d, HD 192310 c, HD 180617 b, and 55 Cnc f), and six have masses larger than 1$M_{Jup}$ (HD 160691 b, 47 UMa b, bet Pic c, HD 10647 b, HD 147513 b, and ups And d).
Out of the 13 HWO-accessible HZ planets, 8\% have $M_p$<5$M_\oplus$, 15\% have 5$M_\oplus$<$M_p$<10$M_\oplus$, 31\% have 10$M_\oplus$<$M_p$<1$M_{Jup}$ and 46\% have $M_p$>1$M_{Jup}$.
This HWO bias towards massive HZ planets happens because the majority of nearby low-mass HZ planets discovered to date orbit M stars at small angular separations.

Some of these HZ planets, in particular those on long-period orbits, might host exomoons potentially detectable through the light modulations produced by mutual transits and shadows \citep{cabrera-schneider2007} or by spectroastrometry \citep{agoletal2015}.
\citet{dobosetal2021} report long-term dynamical stability for a potential exomoon --with survival rates greater than $\sim$\%40-- for nine of the LIFE-detectable HZ planets.
The majority of them are more massive than 1$M_{Jup}$ (HD 160691 b --i.e. mu Arae b--, 47 UMa b, bet Pic c, HD 10647 b, HD 147513 b and ups And d) although some of them have lower masses (HD 40307 g, HD 69830 d, and 55 Cnc f).
Given their rather long orbital periods, we find these nine planets to be common targets for LIFE and reflected-starlight telescopes (Table \ref{table:detectability_overlap_20pc}).
The possible spectral contamination caused by exomoons and the sensitivity required to detect these objects with different instruments will be addressed in future work.

Figure \ref{fig:results_HZ1_2m} shows that most of the LIFE-detectable HZ planets are part of multi-planet systems.
In 20 of these systems, additional planets outside the HZ (both inwards and outwards) are detectable.
LIFE will thus be able to probe a significant number of planets in the regions generally considered too hot or too cold to sustain liquid water on the planetary surface.
Probing the diversity of planetary atmospheres both within and outside the HZ is a key to understanding the requirements for exoplanets to develop potentially habitable environments.

Planet population statistics predict that there should be hundreds of HZ planets within 20~pc from the Sun, most of which are still undiscovered \citep[e.g.][]{quanzetal2022}.
LIFE will be sensitive to most of the newly discovered HZ planets around nearby stars of any spectral type.
Reflected-starlight missions might miss those around M stars, but will be sensitive to those orbiting FGK stars.
Potential Earth analogues around Sun-like stars will therefore be prime targets for synergistic characterization both in thermal emission and reflected starlight.

\section{Discussion} \label{sec:discussion}
In this work we have adopted several assumptions to consistently process and analyse the complete dataset of known exoplanets.
Below we discuss the possible influence of these assumptions on the final results.

\begin{itemize}
    \item We focused on the population of known exoplanets within 20~pc because these are especially favourable for atmospheric characterization. There is however no strict limit on the distances at which LIFE can detect planets. Indeed, we do not find a decrease in the detectability of low-mass planets as $d$ increases within 0-20~pc (Figs. \ref{fig:detectableVSknown_all_missions} and \ref{fig:detectable_vs_known_LIFE_SNR-coloured}). This suggests that 20~pc is a rather conservative limit and extending the analysis further out will increase the target list provided here.
    
    \item We assumed a total 5\% instrument throughput for all LIFE scenarios, which is a rather conservative value. Given that no instrumental errors are considered as yet, our calculations assume photon-noise limit and the yield scales roughly as the square root of the integration time and thus the square-root of the throughput. \citet{kammereretal2022} found that increasing the throughput to 20\% would increase the predicted detection yield by a factor of about two. Such a throughput would reduce the $t_{int}$ presented in this work or, respectively, the mirror size needed to achieve it.  

    \item For all the \texttt{LIFEsim} computations shown in this work we assumed an exozodiacal dust level of 3~zodi. To test the effect of other dust levels, we repeated all computations for a value of 27~zodi \citep[the 95\% confidence upper limit from the HOSTS survey,][]{ertel2020}. We found that the number of LIFE-detectable planets does not change for any of the three scenarios considered. We found nevertheless a mean increase of about 40\% in the required $t_{int}$ to achieve a detection with LIFE's reference configuration (resp. an increase of 20\% and 60\% in $t_{int}$ for LIFE's pessimistic and optimistic configurations).

    \item We assumed all planet-hosting stars as single. While wide binaries will not affect the detectability with LIFE dramatically, close binaries will lead to stellar leakage. This is because with nulling interferometry, only the on-axis source can be suppressed (i.e., nulled) while off-axis sources leak through the interferometer. We note that seven of the systems containing LIFE-detectable HZ planets are known to be multiple. These are: GJ 667 C (triple system), HD 147513 (binary), Proxima Cen (triple), HD 180617 (binary), ups And (binary), HD 147379 (binary), and 55 Cnc (binary).
    
    \item In our $S/N$ computations for LIFE we did not account for the orbital motion of the planets during the integration time. This might blur the signal of the planet and hinder its characterization, especially for short-period planets with long $t_{int}$. On the other hand, accounting for the orbital motion of a planet in multi-epoch imaging campaigns has been shown to help increase the $S/N$ of the planet detection \citep{lecorolleretal2022}. For LIFE's reference configuration, we find six planets with $t_{int}$ greater than 10\% of the orbital period (GJ 367 b, LHS 1678 c, LHS 1140 c, GJ 3929 b, and TRAPPIST-1 b and c). TRAPPIST-1 b and c have indeed $t_{int}$>$P$.
    
    \item Similarly, we did not consider the impact of having several planets in the field of view when observing multi-planet systems \citep[see][for reflected-starlight measurements]{saxena2022}. Accurate ephemerides of the planets in the system will help in extracting the signal of each point source from LIFE's interferometric data. We note that RV alone does not provide information on the orbital inclination, and current astrometry facilities such as Gaia are generally not sensitive to nearby low-mass planets. This highlights additional synergies of LIFE with HWO and with ground-based direct-imaging instruments --which will provide high-accuracy ephemerides. In multi-planet systems, additional integration time may be required to avoid the signals of each planet being contaminated by that of other planets. We note, however, that only five out of the 212 detectable planets with LIFE's reference configuration have $t_{int}$ between 50 and 100 h. Given that exozodi shot noise --already included in \texttt{LIFEsim}-- is expected to dominate over planetary shot noise in the mid-IR, the tally of detectable planets is not expected to be significantly affected. Future work will quantify these effects.

    \item When computing $T_{eq}$ for each planet (Eq. \ref{eq:Teq}) we assumed in all cases $A_B$=0.45 and an efficient heat redistribution consistent with fast-rotating planets ($f$=1). If a planet is tidally locked, the heat redistribution factor $f$ will be 0.5 and the value of $T_{eq}$ will be higher than reported here. 
    
    \item By working with $T_{eq}$, we also assume negligible contribution of internal heat. This effect is especially relevant for young exoplanets, as well as for mature giant ones. It will result in an effective temperature higher than our computed $T_{eq}$. Such an increase in the effective temperature will reduce the $t_{int}$ required to detect the planet. This effect will increase the detectability of mature long-period giant exoplanets --to which HWO is particularly sensitive (Fig. \ref{fig:results_MpvsTeq_2m_onlyHWO})-- in the mid-IR with LIFE. The overlap between both missions is thus expected to be higher than reported in this work. 
    
    \item In Sect. \ref{subsec:results_habitable} we took the habitable zone for a 1$M_\oplus$ planet as defined in \citet{kopparapu2014}. As shown in their work, for planets with $M_p$>1$M_\oplus$ the conservative HZ will actually be wider than the one considered here.
    
    \item To determine whether a planet is in the HZ or not, we took into account the value of the semi-major axis. Depending on the orbital eccentricity, we note that additional planets to those reported in Sect. \ref{subsec:results_habitable} might enter the HZ during a fraction of their orbit. Also, some of the planets in Sect. \ref{subsec:results_habitable} may leave the HZ during a fraction of their orbit depending on their eccentricity. Analysing these effects on the climate of such planets is out of the scope of this work.
    
    \item Our reflected-starlight detectability computations are based only on the angular separation (IWA, OWA) and brightness of the planet ($C_{min}$), assuming $A_g$=0.3 and Lambertian scattering. Additional considerations and mission-specific noise models will be needed to determine if a planet reported here as accessible ($P_{\rm{access}}$>25\%) in reflected starlight is actually detectable.
    
    \item Our approach based on the known exoplanets is limited by the current detection biases and the accuracy of the orbital characterization of these systems. Both factors are expected to improve in the next years with ongoing transit and RV efforts \citep[e.g.][]{quirrenbachetal2014, pepeetal2021, lilloboxetal2022}, astrometric measurements by Gaia \citep[e.g.][]{reyleetal2021} and GRAVITY \citep{lacouretal2020}, and preliminary direct-imaging measurements. 
    
    \item Ground-based ELTs will also directly image known nearby planets \citep{quanzetal2015} and will search for others still unknown \citep{bowensetal2021}. ELTs will complement HWO by directly imaging in reflected starlight a number of nearby low-mass exoplanets around M stars \citep{kasperetal2021}, a challenging population for HWO due to its smaller mirror size. Combined with HWO, ELTs will thus increase the overlap between reflected-starlight facilities and LIFE. We will further analyse these synergies with ELTs in future work.
    
    \item The completeness of our catalogue is limited by the reported parameters in the NASA Exoplanet Archive. Using the Default Parameter Set of the Planetary Systems database ensures self-consistently reported planetary parameters. However, this dataset might lack in some cases parameters that are reported in other references. Also, certain orbital parameters may not be homogeneously reported (see footnote \ref{footnote_argper}).

\end{itemize}

\section{Summary and conclusions} \label{sec:conclusions}
In this work we have shown the potential of a mid-IR space-based nulling interferometer to directly image the thermal emission of the known exoplanets within 20~pc.
The scientific theme of detecting and characterizing temperate terrestrial exoplanets in the mid-IR was given very high priority by ESA's Voyage 2050 Senior Committee report\footnote{\url{https://www.cosmos.esa.int/web/voyage-2050}} and is considered a candidate topic for a future L-class mission in the ESA Science Programme.
We applied our detectability study to several possible configurations of the Large Interferometer For Exoplanets mission concept.
Furthermore, we determined which of these known exoplanets are also accessible with a reflected-starlight mission such as the ${\sim}$6m Habitable Worlds Observatory highlighted in the US Astro 2020 Decadal Survey recommendations, assuming performance estimates motivated by previous studies.

The mid-IR wavelength range is sensitive to multiple gaseous species --e.g. H$_2$O, CO$_2$, CH$_4$ or O$_3$-- relevant for habitability analyses of low-mass exoplanets.
This spectral range is not particularly sensitive to the aerosols in the atmosphere, but this enables to constrain the planetary radius and the atmospheric temperature-pressure profile eventually down to the surface.
On the other hand, the optical and near-IR range will help constrain the properties of the clouds and aerosols in the atmosphere.
These wavelengths can also probe phenomena such as the glint of oceans and the vegetation red-edge.
However, the sensitivity of the visible spectral region to atmospheric aerosols introduces parameter degeneracies when interpreting reflected-starlight measurements, for instance between the optical properties of the aerosols, those of the surface, the abundance of gaseous species and the value of the planet radius if unknown.

Combining observations in thermal emission and in reflected starlight will yield a much more complete characterization of the planet by breaking the parameter degeneracies and uncertainties of each individual technique.
First, multi-wavelength measurements will break the radius-albedo degeneracy, deriving the wavelength-dependent albedo of the planet, which contains key information on the atmospheric composition and cloud structure.
The combination of both spectral ranges will allow us to assess the energy budget of the planet and eventually its surface temperature. 
From that, evidence of atmospheric greenhouse could be inferred if the measured brightness temperature of the planet is higher than its equilibrium temperature.
Planets observable with both techniques will be ideal targets for atmospheric characterization.

We find that the reference configuration of LIFE (four 2-m telescopes with a total throughput of 5\% and wavelength range between 4-18.5 $\mu$m) is able to detect ($S/N$=7 achieved in less than 100~h) 212 of the total 259 (82\%) known exoplanets within 20~pc.
In particular, it can detect 144 planets with $M_{p}$<$M_{Neptune}$ (89\%) and 68 with $M_{p}$>$M_{Neptune}$ (69\%).
The notional HWO mission has a similar performance for massive planets, with 64 nearby exoplanets with $M_{p}$>$M_{Neptune}$ accessible ($P_{\rm{access}}$>25\%) in reflected starlight.
For currently-known low-mass ($M_{p}$<$M_{Neptune}$) planets, the tally drops to 16 for HWO.
This is due to the current observational biases that favour the discovery of low-mass exoplanets around low-mass stars.
Their small planet-star angular separations fall inside the IWA of reflected-starlight instruments, and the planets are therefore masked together with the star (see Fig. \ref{fig:detectableVSknown_all_missions}, bottom row).
On the other hand, a mid-IR nulling interferometer is very effective at detecting these targets.

A total of 55 exoplanets within 20~pc are detectable with LIFE's reference configuration and are also accessible in reflected starlight with HWO.
Most of these common targets are giant planets, although we also find some low-mass ones such as the super-Earths tau Cet e, f and h, HD 20794 d and e, and GJ 514 b. 
LIFE's reference configuration can also detect 163 exoplanets which are not accessible to HWO (Table \ref{table:detectability_onlyLIFE_20pc}).
Of these, 62 planets have masses lower than 5$M_\oplus$, 37 of which require less than 1~h of integration time to be detected with $S/N$=7.
While some overlap already exists between the currently known exoplanets accessible to the assumed LIFE and HWO mission concepts, we expect a much larger overlap between the population of new exoplanets that will be detected by these two missions.
That is because the bulk of currently known exoplanets has been detected by RV and transit techniques at small orbital separations.

Of the 40 exoplanets within 20~pc that have been detected in transit to date, LIFE will be able to detect 32.
These are generally in the super-Earth to Neptune-like mass range with $T_{eq}$ between 500 and 700~K, but also include planets with $T_{eq}$>1000~K.
Eight systems have multiple LIFE-detectable transiting planets.
We find none these 40 transiting planets to be accessible with HWO.
Transiting planets have and will be intensively monitored by current and future instruments --e.g. CHEOPS, JWST, PLATO, Ariel and ground-based telescopes.
Complementary direct-imaging observations could characterize deeper atmospheric layers, although we leave the detailed analysis of multi-technique atmospheric retrievals for future work.

LIFE's reference configuration will detect 38 known exoplanets in the habitable zone of their host star, all of which have been discovered in RV and do not transit.
Of these, 13 planets have masses lower than 5$M_\oplus$ and eight planets have masses between 5$M_\oplus$ and 10$M_\oplus$. 
Even in LIFE's pessimistic configuration with four 1-m telescopes and total 5\% throughput, the tally remains nine and six, respectively.
We find that HWO can access a lower number of known HZ planets (13, of which three have $M_p$<10$M_\oplus$) due to the aforementioned observational biases.

The integration time required for LIFE to characterize the atmospheres of these planets will depend on the planet size and temperature, the distance to the planetary system and the stellar type of the host.
This will determine the $S/N$ and $R$ needed to constrain the abundance of a given species with a certain level of precision, depending on the science case.
Current estimates for LIFE's reference configuration with 5\% throughput range from 1~hour to detect PH$_3$ in the atmosphere of a warm giant planet orbiting a G star at 10~pc \citep{angerhausenetal2022} to between 50 and 100 days to detect CH$_4$ and constrain the abundance of H$_2$O, CO$_2$ and O$_3$ on an Earth twin at 10~pc \citep{konradetal2022, aleietal2022}.
Future work will address this in detail for a number of LIFE-detectable planets.

Characterizing the atmospheres of low-mass temperate exoplanets will represent a giant leap in our understanding of exoplanet diversity.
We have shown that a mid-IR nulling interferometer such as LIFE will provide ground-breaking measurements for a relevant sample of these worlds, while additionally being able to characterize a wider variety of planets spanning several orders of magnitude in age, mass, and stellar irradiation.
LIFE will be able to constrain the planet radius --and hence its density when RV data is available--, the abundance of several potential biomarkers, and the atmospheric T-P profile probing in some cases down to the planet surface.
The target list presented here demonstrates the feasibility of these science goals with the exoplanets already known.
This list is expected to increase significantly in the coming years as a result of long-baseline transit and RV campaigns, thus increasing the potential for analysing a large population of diverse nearby exoplanets.

\begin{acknowledgements}
We thank the anonymous referee for the helpful comments which improved the manuscript. OCG acknowledges the financial support of the API Exoplan\`etes - Observatoire de Paris. Part of work has been carried out within the framework of the National Centre of Competence in Research PlanetS supported by the Swiss National Science Foundation under grants 51NF40\_182901 and 51NF40\_205606. DA, FD, and SPQ acknowledge the financial support of the SNSF.   
This research has made use of the NASA Exoplanet Archive, which is operated by the California Institute of Technology, under contract with the National Aeronautics and Space Administration under the Exoplanet Exploration Program.
\end{acknowledgements}

\bibliographystyle{aa}
\bibliography{references}

\begin{appendix} 

\begin{figure*}
\section{Accessible exoplanets in reflected-starlight beyond 20~pc} \label{sec:appendix_beyond20pc}
        \centering
        \includegraphics[width=8.cm]{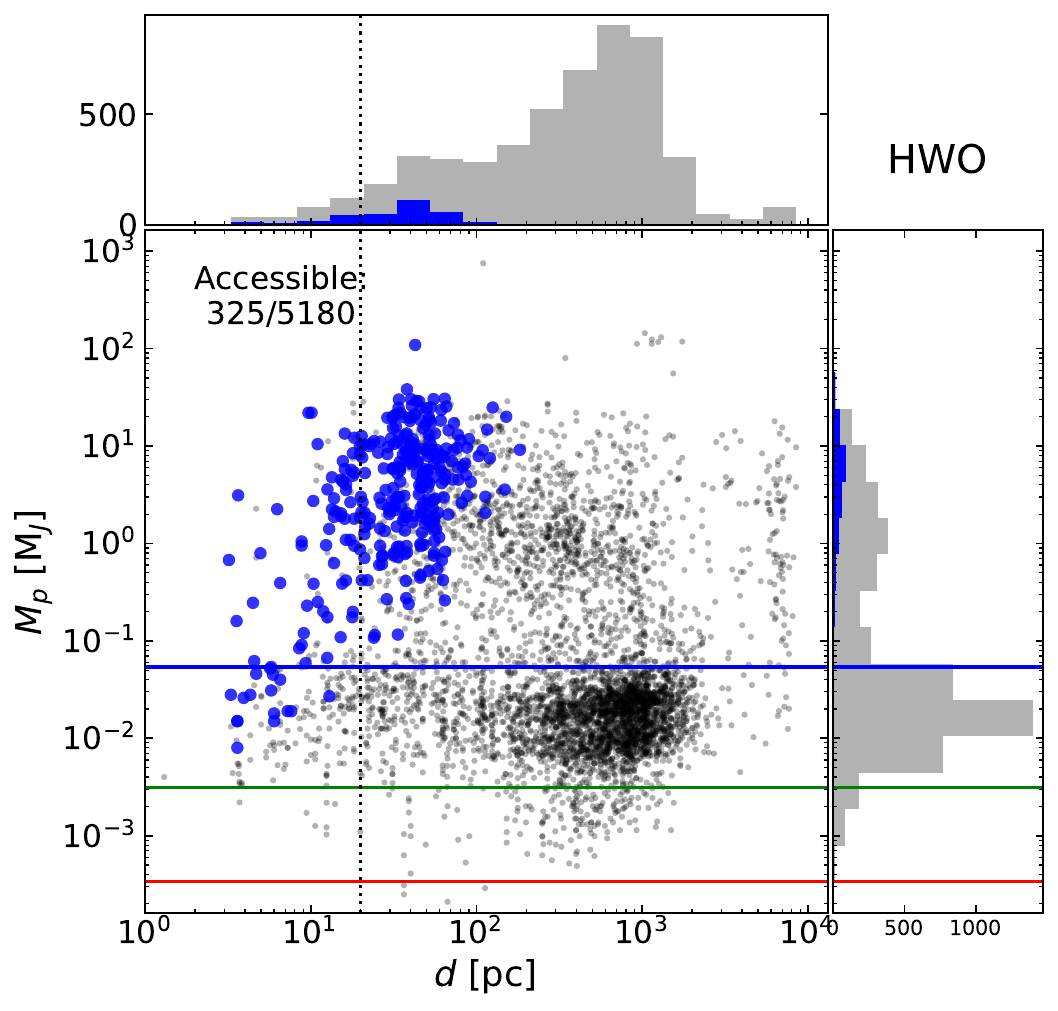} 
        \hfill
        \includegraphics[width=8.cm]{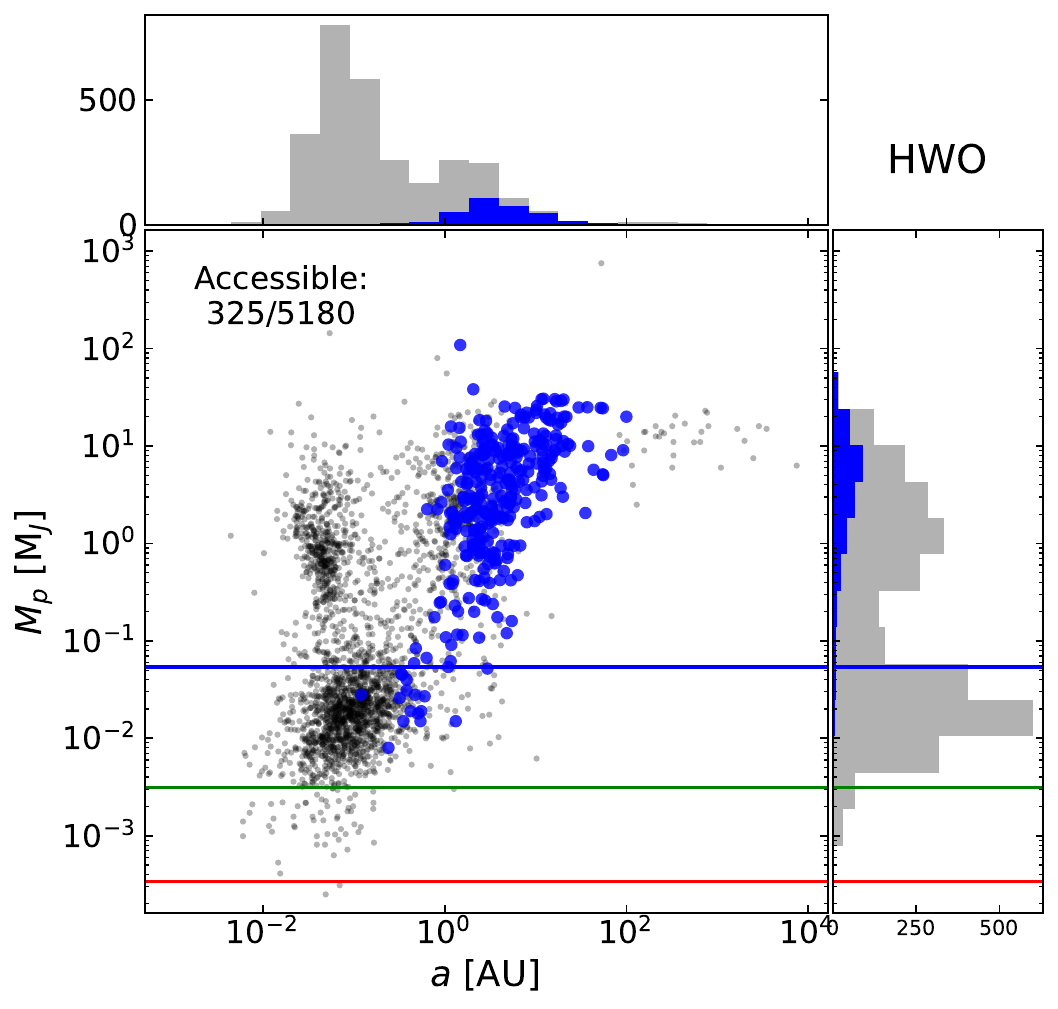}
        \caption{\label{fig:detectableVSknown_all_missions_beyond20pc} As Fig. \ref{fig:detectableVSknown_all_missions}, but including the exoplanets beyond 20~pc with $P_{\rm{access}}$>25\% for the notional HWO mission. The vertical dotted line indicates the 20~pc threshold.}
\end{figure*} 

\begin{figure*}
\section{Exoplanets only accessible with the notional HWO} \label{sec:appendix_onlyHWO}
   \centering
   \includegraphics[width=9.cm]{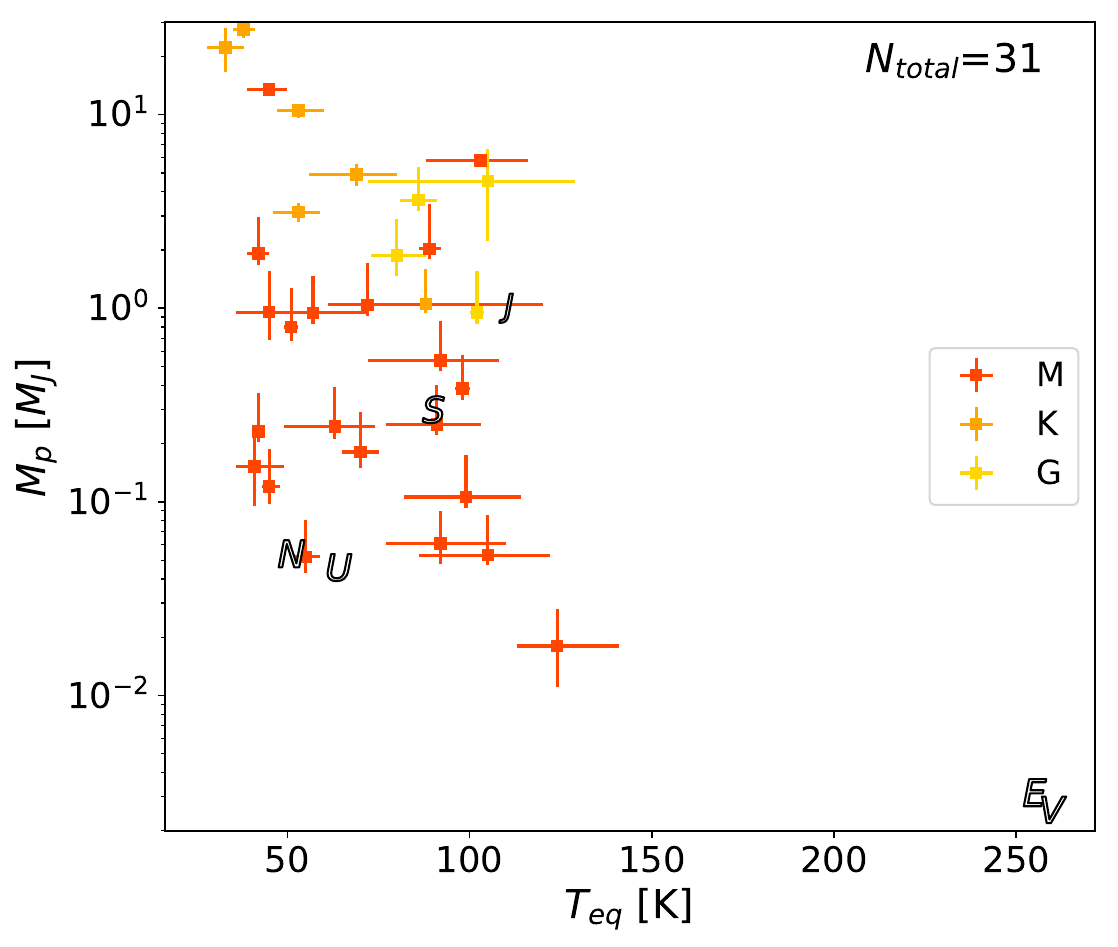}
   \caption{\label{fig:results_MpvsTeq_2m_onlyHWO}
   As Fig. \ref{fig:results_MpvsTeq_2m_onlyLIFE}, but for those planets accessible in reflected starlight with the notional HWO and not with LIFE's reference configuration. These are: GJ 667 C g, HD 115404 A c, GJ 433 c, HD 113538 c, 47 UMa d, HD 153557 d, GJ 317 b, GJ 317 c, GJ 676 A c, GJ 676 A b, GJ 163 d, HIP 38594 c, GJ 229 b, HD 140901 c, HD 145675 c, GJ 3512 c, GJ 3512 b, 55 Cnc d, GJ 687 c, GJ 179 b, GJ 832 b, GJ 9066 c, GJ 680 b, GJ 15 A c, GJ 849 c, HD 95735 c, GJ 1148 c, GJ 849 b, eps Ind A b, TYC 2187-512-1 b, HD 154345 b.}
   \end{figure*}
\FloatBarrier

\begin{figure*}
\section{Integration time of the LIFE-detectable exoplanets.} \label{sec:appendix_LIFEsnr}
   \centering
   \includegraphics[width=9.cm]{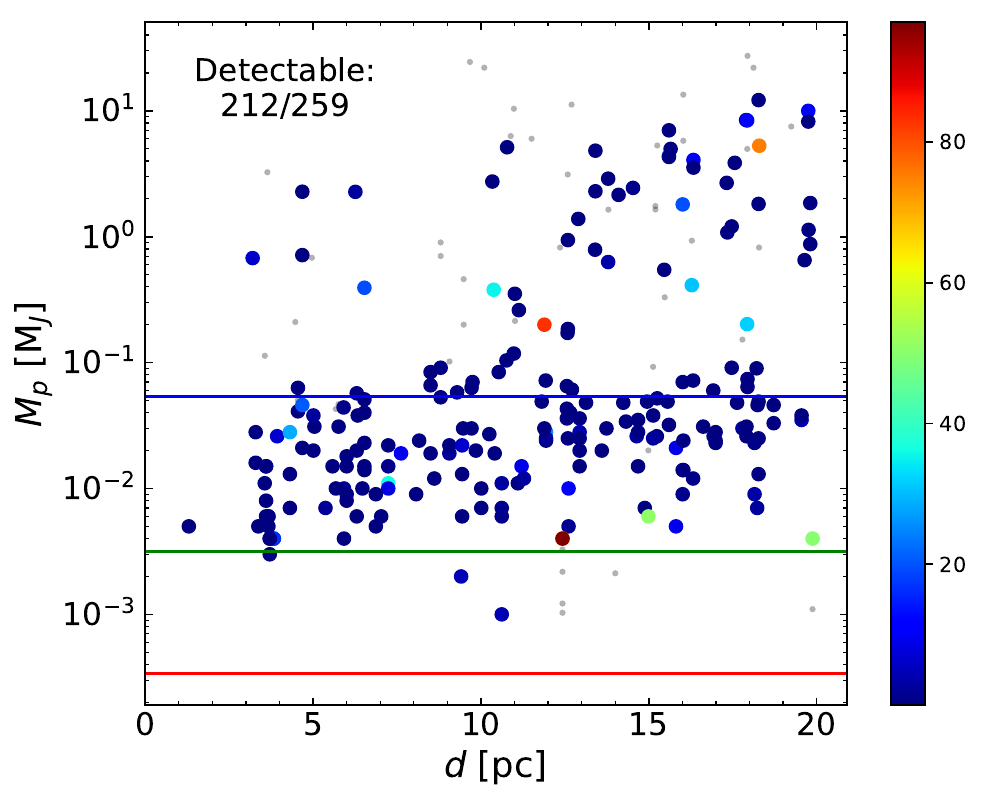}
   \caption{\label{fig:detectable_vs_known_LIFE_SNR-coloured}
   LIFE-detectable exoplanets, as in Fig. \ref{fig:detectableVSknown_all_missions}, but colour-coding each planet according to the integration time (in hours) required to be detected with LIFE's reference configuration.}
   \end{figure*}

\begin{table*}
\section{List of planets only detectable by LIFE} \label{sec:appendix_tableonlyLIFE}
\centering
\small
\caption{Detectability of the exoplanets within 20 pc that are not accessible in reflected starlight with the notional HWO but can be detected ($S/N$=7) with LIFE. Integration times longer than 100 hours are not considered as detectable and are not shown in the table.}
\label{table:detectability_onlyLIFE_20pc} 
\begin{tabular}{l c c c c c c c c c c c}
\hline 
\hline
 & &  &  &  &  &   \multicolumn{2}{c}{LIFE (4x1m)}  & \multicolumn{2}{c}{LIFE (4x2m)}   &  \multicolumn{2}{c}{LIFE (4x3.5m)} \\ 
Planet & $d$ & 	 $a$ & 	 $R_p$  & $M_p$  & $T_{eq}$ & $S/N$ & $t_{\rm{int}}$  & $S/N$ & $t_{\rm{int}}$  & $S/N$ & $t_{\rm{int}}$\\
  & [pc] & [AU] & [$R_{\oplus}$] & [$M_{\oplus}$] & [K] & (10h) & [h] & (10h) & [h] & (10h) & [h] \\
\hline
GJ 367 b & 	 9.41 & 	0.01 & 	0.67 & 	0.64 & 	1178 & 	3 & 	63  & 	10 & 	5  & 	22 & 	1 \\ 
L 98-59 b & 	 10.62 & 	0.02 & 	0.90 & 	0.32 & 	534 & 	3 & 	51  & 	10 & 	4  & 	21 & 	1 \\ 
HD 21749 c & 	 16.32 & 	0.07 & 	0.90 & 	3.81 & 	661 & 	9 & 	6  & 	24 & 	0.8  & 	47 & 	0.2 \\ 
LHS 1678 c & 	 19.88 & 	0.03 & 	1.01 & 	1.27 & 	468 & 	 $-$ & 	 $-$  & 	3 & 	50  & 	7 & 	10 \\ 
YZ Cet b & 	 3.71 & 	0.02 & 	1.01 & 	0.95 & 	410 & 	37 & 	0.4  & 	104 & 	0.04  & 	225 & 	0.01 \\ 
Teegarden's Star c & 	 3.83 & 	0.04 & 	1.12 & 	1.27 & 	191 & 	 $-$ & 	 $-$  & 	5 & 	17  & 	17 & 	2 \\ 
Teegarden's Star b & 	 3.83 & 	0.03 & 	1.12 & 	1.27 & 	253 & 	6 & 	15  & 	21 & 	1  & 	59 & 	0.1 \\ 
GJ 273 c & 	 5.92 & 	0.04 & 	1.12 & 	1.27 & 	401 & 	20 & 	1  & 	56 & 	0.2  & 	123 & 	0.03 \\ 
GJ 1132 b & 	 12.61 & 	0.02 & 	1.12 & 	1.59 & 	516 & 	3 & 	42  & 	17 & 	2  & 	37 & 	0.4 \\ 
GJ 3929 b & 	 15.81 & 	0.03 & 	1.12 & 	1.59 & 	501 & 	 $-$ & 	 $-$  & 	7 & 	9  & 	15 & 	2 \\ 
LTT 1445 A c & 	 6.87 & 	0.03 & 	1.12 & 	1.59 & 	452 & 	20 & 	1  & 	56 & 	0.2  & 	119 & 	0.03 \\ 
LHS 1140 c & 	 14.99 & 	0.03 & 	1.12 & 	1.91 & 	357 & 	 $-$ & 	 $-$  & 	3 & 	51  & 	8 & 	8 \\ 
YZ Cet d & 	 3.71 & 	0.03 & 	1.12 & 	1.27 & 	311 & 	18 & 	1  & 	59 & 	0.1  & 	144 & 	0.02 \\ 
YZ Cet c & 	 3.71 & 	0.02 & 	1.12 & 	1.27 & 	357 & 	22 & 	1  & 	64 & 	0.1  & 	142 & 	0.02 \\ 
TRAPPIST-1 c & 	 12.43 & 	0.016 & 	1.12 & 	1.27 & 	323 & 	 $-$ & 	 $-$  & 	3 & 	73  & 	7 & 	10 \\ 
TRAPPIST-1 b & 	 12.43 & 	0.011 & 	1.12 & 	1.27 & 	379 & 	 $-$ & 	 $-$  & 	2 & 	97  & 	6 & 	15 \\ 
Ross 128 b & 	 3.37 & 	0.05 & 	1.23 & 	1.59 & 	267 & 	13 & 	3  & 	47 & 	0.2  & 	128 & 	0.03 \\ 
GJ 581 e & 	 6.30 & 	0.03 & 	1.23 & 	1.91 & 	485 & 	23 & 	0.9  & 	60 & 	0.1  & 	121 & 	0.03 \\ 
GJ 393 b & 	 7.03 & 	0.05 & 	1.23 & 	1.91 & 	458 & 	26 & 	0.7  & 	72 & 	0.09  & 	156 & 	0.02 \\ 
Proxima Cen b & 	 1.30 & 	0.05 & 	1.23 & 	1.59 & 	229 & 	45 & 	0.2  & 	170 & 	0.02  & 	466 & 	0.002 \\ 
LHS 1478 b & 	 18.23 & 	0.02 & 	1.23 & 	2.22 & 	566 & 	 $-$ & 	 $-$  & 	14 & 	2  & 	31 & 	0.5 \\ 
GJ 357 b & 	 9.44 & 	0.04 & 	1.23 & 	1.91 & 	494 & 	24 & 	0.9  & 	69 & 	0.1  & 	145 & 	0.02 \\ 
GJ 1061 b & 	 3.67 & 	0.02 & 	1.23 & 	1.59 & 	355 & 	38 & 	0.3  & 	113 & 	0.04  & 	257 & 	0.007 \\ 
GJ 1061 d & 	 3.67 & 	0.05 & 	1.23 & 	1.91 & 	248 & 	9 & 	6  & 	38 & 	0.3  & 	111 & 	0.04 \\ 
HD 260655 b & 	 10.01 & 	0.03 & 	1.23 & 	2.22 & 	612 & 	12 & 	3  & 	36 & 	0.4  & 	69 & 	0.1 \\ 
HD 219134 f & 	 6.53 & 	0.15 & 	1.35 & 	7.31 & 	455 & 	19 & 	1  & 	52 & 	0.2  & 	109 & 	0.04 \\ 
tau Cet g & 	 3.60 & 	0.13 & 	1.35 & 	1.91 & 	553 & 	56 & 	0.2  & 	134 & 	0.03  & 	261 & 	0.007 \\ 
LHS 3844 b & 	 14.88 & 	0.01 & 	1.35 & 	2.22 & 	763 & 	4 & 	26  & 	43 & 	0.3  & 	95 & 	0.05 \\ 
Wolf 1061 b & 	 4.31 & 	0.04 & 	1.35 & 	2.22 & 	404 & 	46 & 	0.2  & 	129 & 	0.03  & 	280 & 	0.006 \\ 
LTT 1445 A b & 	 6.87 & 	0.04 & 	1.35 & 	2.86 & 	371 & 	15 & 	2  & 	45 & 	0.2  & 	101 & 	0.05 \\ 
GJ 486 b & 	 8.07 & 	0.02 & 	1.35 & 	2.86 & 	606 & 	18 & 	1  & 	55 & 	0.2  & 	107 & 	0.04 \\ 
L 98-59 c & 	 10.62 & 	0.03 & 	1.35 & 	2.22 & 	451 & 	6 & 	14  & 	19 & 	1  & 	40 & 	0.3 \\ 
HD 20794 c & 	 6.00 & 	0.20 & 	1.35 & 	2.54 & 	477 & 	22 & 	1.0  & 	57 & 	0.1  & 	115 & 	0.04 \\ 
GJ 3323 b & 	 5.37 & 	0.03 & 	1.35 & 	2.22 & 	255 & 	4 & 	29  & 	16 & 	2  & 	43 & 	0.3 \\ 
GJ 1061 c & 	 3.67 & 	0.03 & 	1.35 & 	1.91 & 	274 & 	15 & 	2  & 	49 & 	0.2  & 	124 & 	0.03 \\ 

\hline
\end{tabular}
\end{table*}

\addtocounter{table}{-1}
\begin{table*}[ht]
\centering
\small
\caption{ continued.}
\begin{tabular}{l c c c c c c c c c c c}
\hline 
\hline
 & &  &  &  &  &   \multicolumn{2}{c}{LIFE (4x1m)}  & \multicolumn{2}{c}{LIFE (4x2m)}   &  \multicolumn{2}{c}{LIFE (4x3.5m)} \\ 
Planet & $d$ & 	 $a$ & 	 $R_p$  & $M_p$  & $T_{eq}$ & $S/N$ & $t_{\rm{int}}$  & $S/N$ & $t_{\rm{int}}$  & $S/N$ & $t_{\rm{int}}$\\
  & [pc] & [AU] & [$R_{\oplus}$] & [$M_{\oplus}$] & [K] & (10h) & [h] & (10h) & [h] & (10h) & [h] \\
\hline
HD 219134 c & 	 6.53 & 	0.07 & 	1.46 & 	4.45 & 	676 & 	53 & 	0.2  & 	120 & 	0.03  & 	224 & 	0.01 \\ 
GJ 667 C e & 	 7.24 & 	0.21 & 	1.46 & 	3.50 & 	197 & 	 $-$ & 	 $-$  & 	4 & 	36  & 	11 & 	4 \\ 
GJ 667 C f & 	 7.24 & 	0.15 & 	1.46 & 	3.18 & 	230 & 	 $-$ & 	 $-$  & 	8 & 	8  & 	22 & 	1.0 \\ 
GJ 273 b & 	 5.92 & 	0.09 & 	1.46 & 	3.18 & 	251 & 	5 & 	23  & 	18 & 	2  & 	51 & 	0.2 \\ 
GJ 1132 c & 	 12.61 & 	0.05 & 	1.46 & 	3.18 & 	296 & 	 $-$ & 	 $-$  & 	6 & 	12  & 	17 & 	2 \\ 
GJ 625 b & 	 6.47 & 	0.08 & 	1.46 & 	3.18 & 	292 & 	9 & 	6  & 	33 & 	0.4  & 	87 & 	0.06 \\ 
G 264-012 b & 	 16.01 & 	0.02 & 	1.46 & 	2.86 & 	552 & 	4 & 	26  & 	21 & 	1  & 	43 & 	0.3 \\ 
HD 20794 b & 	 6.00 & 	0.12 & 	1.46 & 	2.86 & 	619 & 	35 & 	0.4  & 	85 & 	0.07  & 	165 & 	0.02 \\ 
GJ 3998 b & 	 18.15 & 	0.03 & 	1.46 & 	2.86 & 	637 & 	3 & 	45  & 	12 & 	3  & 	25 & 	0.8 \\ 
GJ 411 b & 	 5.68 & 	0.08 & 	1.46 & 	3.18 & 	335 & 	17 & 	2  & 	56 & 	0.2  & 	135 & 	0.03 \\ 
HD 219134 d & 	 6.53 & 	0.24 & 	1.57 & 	16.21 & 	357 & 	16 & 	2  & 	48 & 	0.2  & 	109 & 	0.04 \\ 
HD 219134 b & 	 6.53 & 	0.04 & 	1.57 & 	4.77 & 	876 & 	67 & 	0.1  & 	154 & 	0.02  & 	278 & 	0.006 \\ 
GJ 740 b & 	 11.10 & 	0.03 & 	1.57 & 	3.50 & 	730 & 	25 & 	0.8  & 	70 & 	0.1  & 	132 & 	0.03 \\ 
L 98-59 e & 	 10.62 & 	0.07 & 	1.57 & 	3.50 & 	294 & 	4 & 	35  & 	13 & 	3  & 	34 & 	0.4 \\ 
L 98-59 d & 	 10.62 & 	0.05 & 	1.57 & 	1.91 & 	356 & 	7 & 	11  & 	21 & 	1  & 	50 & 	0.2 \\ 
HD 260655 c & 	 10.01 & 	0.05 & 	1.57 & 	3.18 & 	481 & 	19 & 	1  & 	51 & 	0.2  & 	105 & 	0.04 \\ 
GJ 15 A b & 	 3.56 & 	0.07 & 	1.68 & 	3.50 & 	347 & 	49 & 	0.2  & 	148 & 	0.02  & 	339 & 	0.004 \\ 
CD Cet b & 	 8.61 & 	0.02 & 	1.68 & 	3.81 & 	434 & 	20 & 	1  & 	79 & 	0.08  & 	175 & 	0.02 \\ 
HD 136352 b & 	 14.68 & 	0.10 & 	1.68 & 	4.77 & 	780 & 	27 & 	0.7  & 	65 & 	0.1  & 	121 & 	0.03 \\ 
HD 85512 b & 	 11.28 & 	0.26 & 	1.68 & 	3.81 & 	296 & 	4 & 	35  & 	13 & 	3  & 	35 & 	0.4 \\ 
pi Men c & 	 18.27 & 	0.07 & 	1.79 & 	4.13 & 	1048 & 	22 & 	1  & 	61 & 	0.1  & 	115 & 	0.04 \\ 
Wolf 1061 c & 	 4.31 & 	0.09 & 	1.79 & 	4.13 & 	261 & 	14 & 	2  & 	53 & 	0.2  & 	145 & 	0.02 \\ 
GJ 357 c & 	 9.44 & 	0.06 & 	1.79 & 	4.13 & 	378 & 	22 & 	1  & 	69 & 	0.1  & 	165 & 	0.02 \\ 
GJ 667 C c & 	 7.24 & 	0.12 & 	1.91 & 	4.77 & 	256 & 	5 & 	17  & 	20 & 	1  & 	57 & 	0.1 \\ 
Ross 508 b & 	 11.21 & 	0.05 & 	1.91 & 	4.77 & 	266 & 	 $-$ & 	 $-$  & 	7 & 	9  & 	21 & 	1 \\ 
GJ 251 b & 	 5.58 & 	0.08 & 	1.91 & 	4.77 & 	302 & 	19 & 	1  & 	65 & 	0.1  & 	162 & 	0.02 \\ 
G 264-012 c & 	 16.01 & 	0.05 & 	1.91 & 	4.45 & 	358 & 	5 & 	23  & 	18 & 	2  & 	43 & 	0.3 \\ 
GJ 676 A d & 	 16.03 & 	0.04 & 	1.91 & 	4.45 & 	689 & 	15 & 	2  & 	44 & 	0.3  & 	85 & 	0.07 \\ 
55 Cnc e & 	 12.59 & 	0.02 & 	1.91 & 	7.95 & 	1678 & 	21 & 	1  & 	75 & 	0.09  & 	172 & 	0.02 \\ 
GJ 887 b & 	 3.29 & 	0.07 & 	2.02 & 	5.09 & 	404 & 	97 & 	0.05  & 	265 & 	0.007  & 	560 & 	0.002 \\ 
HD 40307 b & 	 12.94 & 	0.05 & 	2.02 & 	4.77 & 	818 & 	56 & 	0.2  & 	150 & 	0.02  & 	278 & 	0.006 \\ 
GJ 433 d & 	 9.06 & 	0.18 & 	2.24 & 	6.04 & 	234 & 	3 & 	41  & 	14 & 	3  & 	41 & 	0.3 \\ 
GJ 682 b & 	 5.01 & 	0.08 & 	2.24 & 	6.36 & 	238 & 	12 & 	4  & 	45 & 	0.2  & 	124 & 	0.03 \\ 
61 Vir b & 	 8.50 & 	0.05 & 	2.24 & 	6.04 & 	1017 & 	108 & 	0.04  & 	257 & 	0.007  & 	463 & 	0.002 \\ 
HD 40307 f & 	 12.94 & 	0.25 & 	2.24 & 	6.36 & 	354 & 	10 & 	5  & 	35 & 	0.4  & 	87 & 	0.07 \\ 
GJ 581 c & 	 6.30 & 	0.07 & 	2.35 & 	6.36 & 	302 & 	25 & 	0.8  & 	81 & 	0.07  & 	200 & 	0.01 \\ 
Gl 49 b & 	 9.85 & 	0.09 & 	2.35 & 	6.36 & 	402 & 	37 & 	0.4  & 	110 & 	0.04  & 	250 & 	0.008 \\ 
HD 22496 b & 	 13.60 & 	0.05 & 	2.35 & 	6.36 & 	729 & 	84 & 	0.07  & 	221 & 	0.01  & 	419 & 	0.003 \\ 
GJ 876 d & 	 4.68 & 	0.02 & 	2.35 & 	6.67 & 	531 & 	159 & 	0.02  & 	403 & 	0.003  & 	777 & 	0.0008 \\ 
GJ 536 b & 	 10.41 & 	0.07 & 	2.35 & 	6.04 & 	420 & 	29 & 	0.6  & 	81 & 	0.07  & 	173 & 	0.02 \\ 
GJ 667 C b & 	 7.24 & 	0.05 & 	2.47 & 	6.99 & 	406 & 	56 & 	0.2  & 	153 & 	0.02  & 	327 & 	0.005 \\ 
GJ 433 b & 	 9.06 & 	0.06 & 	2.47 & 	6.99 & 	408 & 	34 & 	0.4  & 	93 & 	0.06  & 	197 & 	0.01 \\ 
GJ 3929 c & 	 15.81 & 	0.08 & 	2.47 & 	6.67 & 	279 & 	 $-$ & 	 $-$  & 	7 & 	9  & 	20 & 	1 \\ 
GJ 357 d & 	 9.44 & 	0.20 & 	2.47 & 	6.99 & 	205 & 	 $-$ & 	 $-$  & 	9 & 	6  & 	28 & 	0.6 \\ 
GJ 180 c & 	 11.94 & 	0.12 & 	2.58 & 	7.95 & 	257 & 	5 & 	24  & 	17 & 	2  & 	47 & 	0.2 \\ 
GJ 143 b & 	 16.32 & 	0.19 & 	2.58 & 	22.88 & 	372 & 	13 & 	3  & 	42 & 	0.3  & 	100 & 	0.05 \\ 
HD 7924 b & 	 16.99 & 	0.06 & 	2.58 & 	7.31 & 	769 & 	43 & 	0.3  & 	121 & 	0.03  & 	227 & 	0.009 \\ 
HD 136352 d & 	 14.68 & 	0.43 & 	2.58 & 	8.90 & 	372 & 	11 & 	4  & 	34 & 	0.4  & 	81 & 	0.07 \\ 
GJ 3998 c & 	 18.15 & 	0.09 & 	2.58 & 	7.31 & 	362 & 	7 & 	11  & 	22 & 	1  & 	52 & 	0.2 \\ 
GJ 180 b & 	 11.94 & 	0.09 & 	2.69 & 	7.63 & 	296 & 	8 & 	7  & 	27 & 	0.7  & 	69 & 	0.1 \\ 
Gl 686 b & 	 8.16 & 	0.09 & 	2.69 & 	7.63 & 	330 & 	31 & 	0.5  & 	99 & 	0.05  & 	236 & 	0.009 \\ 
HD 7924 d & 	 16.99 & 	0.15 & 	2.69 & 	7.63 & 	482 & 	32 & 	0.5  & 	90 & 	0.06  & 	194 & 	0.01 \\ 
HD 154088 b & 	 18.27 & 	0.13 & 	2.69 & 	7.95 & 	633 & 	51 & 	0.2  & 	133 & 	0.03  & 	267 & 	0.007 \\ 
HD 40307 c & 	 12.94 & 	0.08 & 	2.69 & 	7.95 & 	620 & 	73 & 	0.09  & 	181 & 	0.01  & 	346 & 	0.004 \\ 
GJ 676 A e & 	 16.03 & 	0.18 & 	2.69 & 	7.63 & 	320 & 	8 & 	9  & 	26 & 	0.7  & 	66 & 	0.1 \\ 
GJ 163 c & 	 15.13 & 	0.13 & 	2.69 & 	7.95 & 	264 & 	4 & 	40  & 	13 & 	3  & 	36 & 	0.4 \\ 
GJ 1214 b & 	 14.64 & 	0.01 & 	2.69 & 	8.26 & 	508 & 	 $-$ & 	 $-$  & 	12 & 	4  & 	25 & 	0.8 \\ 
GJ 3942 b & 	 16.93 & 	0.06 & 	2.80 & 	8.26 & 	514 & 	21 & 	1  & 	61 & 	0.1  & 	122 & 	0.03 \\ 
GJ 1265 b & 	 10.25 & 	0.03 & 	2.80 & 	8.58 & 	364 & 	12 & 	4  & 	46 & 	0.2  & 	109 & 	0.04 \\ 
HD 189567 c & 	 17.91 & 	0.20 & 	2.80 & 	8.26 & 	552 & 	36 & 	0.4  & 	97 & 	0.05  & 	202 & 	0.01 \\ 
HD 238090 b & 	 15.24 & 	0.09 & 	2.80 & 	8.26 & 	420 & 	24 & 	0.9  & 	69 & 	0.1  & 	151 & 	0.02 \\ 
GJ 180 d & 	 11.94 & 	0.31 & 	2.91 & 	8.90 & 	184 & 	 $-$ & 	 $-$  & 	4 & 	26  & 	14 & 	2 \\ 
AU Mic c & 	 9.72 & 	0.11 & 	2.91 & 	9.53 & 	444 & 	51 & 	0.2  & 	154 & 	0.02  & 	353 & 	0.004 \\ 
HD 7924 c & 	 16.99 & 	0.11 & 	2.91 & 	8.90 & 	555 & 	49 & 	0.2  & 	127 & 	0.03  & 	252 & 	0.008 \\ 

\hline
\end{tabular}
\end{table*}

\addtocounter{table}{-1}
\begin{table*}[ht]
\centering
\small
\caption{ continued.}
\begin{tabular}{l c c c c c c c c c c c}
\hline 
\hline
 & &  &  &  &  &   \multicolumn{2}{c}{LIFE (4x1m)}  & \multicolumn{2}{c}{LIFE (4x2m)}   &  \multicolumn{2}{c}{LIFE (4x3.5m)} \\ 
Planet & $d$ & 	 $a$ & 	 $R_p$  & $M_p$  & $T_{eq}$ & $S/N$ & $t_{\rm{int}}$  & $S/N$ & $t_{\rm{int}}$  & $S/N$ & $t_{\rm{int}}$\\
  & [pc] & [AU] & [$R_{\oplus}$] & [$M_{\oplus}$] & [K] & (10h) & [h] & (10h) & [h] & (10h) & [h] \\
\hline

GJ 414 A b & 	 11.89 & 	0.23 & 	2.91 & 	9.53 & 	319 & 	12 & 	3  & 	44 & 	0.3  & 	114 & 	0.04 \\ 
HD 136352 c & 	 14.68 & 	0.17 & 	2.91 & 	11.12 & 	584 & 	52 & 	0.2  & 	132 & 	0.03  & 	263 & 	0.007 \\ 
HD 160691 d & 	 15.60 & 	0.09 & 	3.03 & 	10.17 & 	995 & 	123 & 	0.03  & 	286 & 	0.006  & 	524 & 	0.002 \\ 
HIP 38594 b & 	 17.79 & 	0.26 & 	3.03 & 	9.53 & 	270 & 	4 & 	36  & 	14 & 	2  & 	41 & 	0.3 \\ 
HD 285968 b & 	 9.47 & 	0.07 & 	3.03 & 	9.53 & 	415 & 	56 & 	0.2  & 	155 & 	0.02  & 	331 & 	0.004 \\ 
GJ 3779 b & 	 13.74 & 	0.03 & 	3.03 & 	9.53 & 	454 & 	15 & 	2  & 	59 & 	0.1  & 	127 & 	0.03 \\ 
GJ 3082 b & 	 16.62 & 	0.08 & 	3.03 & 	9.85 & 	405 & 	19 & 	1  & 	60 & 	0.1  & 	134 & 	0.03 \\ 
HD 26965 b & 	 5.04 & 	0.22 & 	3.14 & 	9.85 & 	419 & 	120 & 	0.03  & 	323 & 	0.005  & 	670 & 	0.001 \\ 
HD 189567 b & 	 17.91 & 	0.11 & 	3.14 & 	9.85 & 	741 & 	83 & 	0.07  & 	204 & 	0.01  & 	383 & 	0.003 \\ 
HIP 54373 b & 	 18.72 & 	0.06 & 	3.14 & 	10.49 & 	509 & 	20 & 	1  & 	62 & 	0.1  & 	126 & 	0.03 \\ 
GJ 685 b & 	 14.32 & 	0.13 & 	3.25 & 	10.81 & 	317 & 	13 & 	3  & 	44 & 	0.3  & 	107 & 	0.04 \\ 
HD 216520 c & 	 19.55 & 	0.53 & 	3.25 & 	11.12 & 	256 & 	3 & 	65  & 	11 & 	4  & 	31 & 	0.5 \\ 
GJ 682 c & 	 5.01 & 	0.19 & 	3.36 & 	12.08 & 	161 & 	3 & 	53  & 	14 & 	3  & 	46 & 	0.2 \\ 
HD 40307 d & 	 12.94 & 	0.13 & 	3.36 & 	11.44 & 	482 & 	72 & 	0.09  & 	197 & 	0.01  & 	413 & 	0.003 \\ 
HD 69830 b & 	 12.56 & 	0.08 & 	3.47 & 	11.44 & 	757 & 	159 & 	0.02  & 	379 & 	0.003  & 	702 & 	0.001 \\ 
HD 216520 b & 	 19.55 & 	0.20 & 	3.47 & 	12.08 & 	417 & 	27 & 	0.7  & 	82 & 	0.07  & 	186 & 	0.01 \\ 
GJ 338 B b & 	 6.33 & 	0.14 & 	3.59 & 	12.08 & 	339 & 	72 & 	0.09  & 	230 & 	0.009  & 	546 & 	0.002 \\
GJ 163 b & 	 15.13 & 	0.06 & 	3.59 & 	12.08 & 	379 & 	17 & 	2  & 	54 & 	0.2  & 	123 & 	0.03 \\ 
GJ 422 b & 	 12.66 & 	0.11 & 	3.70 & 	12.71 & 	253 & 	7 & 	10  & 	26 & 	0.7  & 	71 & 	0.1 \\ 
GJ 674 b & 	 4.55 & 	0.04 & 	3.70 & 	13.03 & 	464 & 	429 & 	0.003  & 	1093 & 	0.0004  & 	2195 & 	0.0001 \\ 
HD 69830 c & 	 12.56 & 	0.19 & 	3.81 & 	13.67 & 	493 & 	73 & 	0.09  & 	209 & 	0.01  & 	454 & 	0.002 \\ 
GJ 3222 b & 	 18.24 & 	0.09 & 	3.81 & 	14.62 & 	1671 & 	340 & 	0.004  & 	1180 & 	0.0003  & 	2441 & 	8e-05 \\ 
HD 211970 b & 	 13.13 & 	0.14 & 	4.04 & 	15.26 & 	348 & 	36 & 	0.4  & 	114 & 	0.04  & 	271 & 	0.007 \\ 
LSPM J21+02 b\tablefoottext{a} & 	 17.63 & 	0.09 & 	4.04 & 	15.26 & 	325 & 	12 & 	3  & 	41 & 	0.3  & 	102 & 	0.05 \\ 
HIP 54373 c & 	 18.72 & 	0.10 & 	4.04 & 	14.62 & 	407 & 	29 & 	0.6  & 	88 & 	0.06  & 	197 & 	0.01 \\ 
GJ 436 b & 	 9.75 & 	0.03 & 	4.15 & 	22.25 & 	598 & 	152 & 	0.02  & 	419 & 	0.003  & 	804 & 	0.0008 \\ 
GJ 480 b & 	 14.24 & 	0.07 & 	4.15 & 	15.26 & 	367 & 	27 & 	0.7  & 	85 & 	0.07  & 	193 & 	0.01 \\ 
HD 141004 b & 	 11.81 & 	0.12 & 	4.15 & 	15.57 & 	828 & 	186 & 	0.01  & 	420 & 	0.003  & 	778 & 	0.0008 \\ 
GJ 720 A b & 	 15.56 & 	0.12 & 	4.15 & 	15.57 & 	347 & 	27 & 	0.7  & 	83 & 	0.07  & 	196 & 	0.01 \\ 
pi Men d & 	 18.27 & 	0.50 & 	4.15 & 	15.57 & 	393 & 	25 & 	0.8  & 	77 & 	0.08  & 	178 & 	0.02 \\ 
Gl 378 b & 	 14.95 & 	0.04 & 	4.15 & 	15.57 & 	611 & 	72 & 	0.09  & 	215 & 	0.01  & 	418 & 	0.003 \\ 
HD 192310 b & 	 8.80 & 	0.32 & 	4.26 & 	16.84 & 	355 & 	61 & 	0.1  & 	193 & 	0.01  & 	454 & 	0.002 \\ 
HD 140901 b & 	 15.25 & 	0.08 & 	4.37 & 	16.53 & 	867 & 	197 & 	0.01  & 	511 & 	0.002  & 	941 & 	0.0006 \\ 
HD 177565 b & 	 16.92 & 	0.25 & 	4.48 & 	19.07 & 	467 & 	54 & 	0.2  & 	161 & 	0.02  & 	364 & 	0.004 \\ 
GJ 581 b & 	 6.30 & 	0.04 & 	4.60 & 	18.12 & 	402 & 	211 & 	0.01  & 	567 & 	0.002  & 	1181 & 	0.0003 \\ 
AU Mic b & 	 9.72 & 	0.06 & 	4.71 & 	20.02 & 	500 & 	141 & 	0.02  & 	347 & 	0.004  & 	676 & 	0.001 \\ 
HD 190007 b & 	 12.71 & 	0.09 & 	4.71 & 	19.39 & 	565 & 	166 & 	0.02  & 	411 & 	0.003  & 	798 & 	0.0008 \\ 
HD 153557 c & 	 17.94 & 	0.11 & 	4.93 & 	20.34 & 	608 & 	207 & 	0.01  & 	545 & 	0.002  & 	1088 & 	0.0004 \\ 
GJ 687 b & 	 4.55 & 	0.16 & 	4.93 & 	20.02 & 	231 & 	63 & 	0.1  & 	239 & 	0.009  & 	645 & 	0.001 \\ 
61 Vir c & 	 8.50 & 	0.22 & 	5.04 & 	20.98 & 	491 & 	198 & 	0.01  & 	540 & 	0.002  & 	1131 & 	0.0004 \\ 
HD 190360 c & 	 16.01 & 	0.13 & 	5.16 & 	22.25 & 	713 & 	225 & 	0.01  & 	536 & 	0.002  & 	1018 & 	0.0005 \\ 
GJ 96 b & 	 11.93 & 	0.29 & 	5.27 & 	22.88 & 	245 & 	14 & 	2  & 	57 & 	0.2  & 	168 & 	0.02 \\ 
HD 153557 b & 	 17.94 & 	0.07 & 	5.38 & 	23.52 & 	669 & 	149 & 	0.02  & 	413 & 	0.003  & 	784 & 	0.0008 \\ 
HIP 48714 b & 	 10.53 & 	0.11 & 	5.72 & 	26.70 & 	444 & 	229 & 	0.009  & 	682 & 	0.001  & 	1524 & 	0.0002 \\ 
HD 99492 b & 	 18.21 & 	0.12 & 	5.72 & 	28.60 & 	540 & 	158 & 	0.02  & 	419 & 	0.003  & 	852 & 	0.0007 \\ 
rho CrB c & 	 17.47 & 	0.41 & 	6.16 & 	28.92 & 	425 & 	65 & 	0.1  & 	192 & 	0.01  & 	427 & 	0.003 \\ 
HD 147379 b & 	 10.76 & 	0.32 & 	6.73 & 	33.05 & 	242 & 	28 & 	0.6  & 	110 & 	0.04  & 	308 & 	0.005 \\ 
HD 115404 A b & 	 10.98 & 	0.09 & 	7.17 & 	37.50 & 	618 & 	645 & 	0.001  & 	1547 & 	0.0002  & 	2945 & 	6e-05 \\ 
55 Cnc c & 	 12.59 & 	0.24 & 	9.64 & 	58.80 & 	428 & 	270 & 	0.007  & 	795 & 	0.0008  & 	1763 & 	0.0002 \\ 
70 Vir b & 	 17.90 & 	0.48 & 	11.99 & 	2683.64 & 	493 & 	307 & 	0.005  & 	852 & 	0.0007  & 	1794 & 	0.0001 \\ 
HD 3651 b & 	 11.13 & 	0.29 & 	12.11 & 	82.95 & 	487 & 	858 & 	0.0007  & 	2362 & 	9e-05  & 	4912 & 	2e-05 \\ 
tau Boo b & 	 15.65 & 	0.05 & 	12.33 & 	1582.73 & 	1454 & 	937 & 	0.0006  & 	2922 & 	6e-05  & 	5933 & 	1e-05 \\ 
GJ 86 b & 	 10.78 & 	0.11 & 	12.33 & 	1631.35 & 	570 & 	1491 & 	0.0002  & 	3507 & 	4e-05  & 	6617 & 	1e-05 \\ 
GJ 3021 b & 	 17.56 & 	0.49 & 	12.44 & 	1226.77 & 	362 & 	198 & 	0.01  & 	646 & 	0.001  & 	1532 & 	0.0002 \\ 
HD 189733 b & 	 19.76 & 	0.03 & 	12.67 & 	360.09 & 	1035 & 	369 & 	0.004  & 	1331 & 	0.0003  & 	2756 & 	6e-05 \\ 
HIP 79431 b & 	 14.53 & 	0.36 & 	12.78 & 	776.11 & 	156 & 	4 & 	30  & 	19 & 	1  & 	63 & 	0.1 \\ 
GJ 876 b & 	 4.68 & 	0.21 & 	12.89 & 	723.35 & 	164 & 	54 & 	0.2  & 	241 & 	0.008  & 	761 & 	0.0008 \\ 
rho CrB b & 	 17.47 & 	0.22 & 	13.23 & 	383.29 & 	582 & 	676 & 	0.001  & 	1653 & 	0.0002  & 	3199 & 	5e-05 \\ 
55 Cnc b & 	 12.59 & 	0.11 & 	13.34 & 	298.75 & 	619 & 	1363 & 	0.0003  & 	3176 & 	5e-05  & 	5937 & 	1e-05 \\ 
ups And b & 	 13.40 & 	0.06 & 	13.45 & 	250.44 & 	1313 & 	1481 & 	0.0002  & 	4038 & 	3e-05  & 	7558 & 	1e-05 \\ 
HD 192263 b & 	 19.64 & 	0.15 & 	13.67 & 	206.90 & 	459 & 	471 & 	0.002  & 	1263 & 	0.0003  & 	2567 & 	7e-05 \\ 
GJ 876 c & 	 4.68 & 	0.13 & 	13.67 & 	226.92 & 	215 & 	277 & 	0.006  & 	1061 & 	0.0004  & 	2816 & 	6e-05 \\ 
51 Peg b & 	 15.46 & 	0.05 & 	13.79 & 	173.53 & 	1139 & 	1096 & 	0.0004  & 	3257 & 	5e-05  & 	6332 & 	1e-05 \\ 
GJ 1148 b & 	 11.01 & 	0.17 & 	14.01 & 	111.87 & 	217 & 	67 & 	0.1  & 	271 & 	0.007  & 	790 & 	0.0008 \\ 

\hline
\end{tabular}
\newline
\tablefoottext{a}{LSPM J2116+0234 b}
\end{table*}
\FloatBarrier

\section{Science cases of low-mass ($M_p$<5$M_\oplus$) habitable-zone planets detectable by LIFE} \label{sec:appendix_HZ_science_cases}

\begin{description}
    \item[Proxima Cen b:] Located in the conservative HZ of our closest stellar neighbour, it has a minimum mass of 1.27$M_\oplus$ and an eccentricity $e$<0.35 \citep{angladaescudeetal2016}.
    The M5.5V star has shown high activity levels, with XUV irradiance 60 times that of the Earth \citep{ribasetal2017} and frequent flaring events \citep{macgregoretal2018, howardetal2018, vidaetal2019}.
    This stellar activity will affect the habitability conditions of Proxima b \citep{scheucheretal2020} and might lead to atmospheric erosion depending on the planetary magnetic field \citep{dongetal2017}.
    Although potential transits have been reported \citep{kippingetal2017, liuetal2018}, no robust evidence has been found as these photometric variations can be caused by stellar activity \citep{felizetal2019, jenkinsetal2019}.
    
    \citet{lovisetal2017} found that Proxima Cen b could be detected in 20--40 nights of telescope time by coupling upgraded versions of ESPRESSO and SPHERE, with a detection of atmospheric O$_2$ achievable in 60 nights spread over three years.
    JWST's MIRI has been proposed to detect the eventual atmosphere of the planet either by measuring thermal phase curves \citep{kreidberg-loeb2016} or through cross-correlation \citep{snellenetal2017}.
    The unknown planet-to-star ratio, the stellar activity and the lack of a measurement of the orbital inclination could however prevent the interpretation of these observations.
    We find that only LIFE ($S/N$=170 in 10~h of integration time) will be able to directly image Proxima Cen b (Table \ref{table:detectability_overlap_20pc}).
    LIFE will probe the atmospheric components of the planet \citep{defrereetal2018} and its T-P profile, but also its mass and radius.
    This is key to understand the possible history of the planet, including geological processes such as outgassing which shape the evolution of the atmosphere \citep{noacketal2021}.

    \item[Ross 128 b:] This 1.5$M_\oplus$ planet orbits at the inner edge of the optimistic HZ of one of the closest ($d$=3.3~pc) known planetary systems \citep{bonfilsetal2018b}.
    The M4 host star shows weak magnetic activity and K2 light curves yielded no transits of the planet \citep{bonfilsetal2018}.
    Based on the stellar properties, \citet{soutoetal2018} and \citet{herathetal2021} suggested a rocky composition with an iron core proportionally larger than that of the Earth.
    
    \item[tau Cet e:] A planetary system of four super-Earths with minimum masses between 1.7--3.9$M_\oplus$ was detected in RV around this Sun-like (G8.5V) star \citep{fengetal2017}.
    The age of the system shown in Fig. \ref{fig:results_HZ1_2m} corresponds to the lower limit reported in the Composite database in the NASA Archive from \citet{takedaetal2007}, but we also note that \citet{mamajek-hillenbrand2008} estimate an age of 5.8~Gyr for this star.
    We find a mass of about 4.8$M_\oplus$ for tau Cet e, which orbits in the inner region of the optimistic HZ with $e\sim$0.2.
    Planet f, with similar mass, also orbits near --but slightly outside-- the outer boundary of the HZ.
    Planets e and f are detectable with LIFE and also accessible to HWO in reflected starlight (Table \ref{table:detectability_overlap_20pc}).
    If additional planets are present in the HZ \citep{dietrich-apai2021}, they will fall within the detectability region as well.
    The presence of a debris-disk \citep{greavesetal2004, lawleretal2014} with an inner edge at around 6~AU \citep{macgregoretal2016, hunzikeretal2020} might affect the detectability of the planets depending on its inclination and exozodi abundance.
    
    \item[GJ 1061 d:] This rather inactive M5.5V star hosts three rocky planets (b, c, d) with minimum masses of 1.37, 1.74 and 1.64$M_\oplus$ \citep{dreizleretal2020}. Planet d orbits within the conservative HZ. Planet c, with $T_{eq}$=274~K, orbits close to the inner edge of the optimistic HZ. Upper limits of 0.29 and 0.53 were found for the eccentricity of planets c and d. Follow-up observations will help determine whether GJ 1061 d is one of the few highly eccentric HZ rocky planets known to date and the potential implications for its climate.

    \item[GJ 273 b:] \citet{astudillodefru2017} found two planet signals (GJ 273 b and c) in HARPS RV data, with two additional candidates in the sub-Neptune regime.
    Based on dynamical studies, \citet{pozuelosetal2020} found a mass of 2.89$M_\oplus$<$M_p$<3.03$M_\oplus$ for planet b.
    We find a slightly higher $M_p$=3.2$M_\oplus$ and $T_{eq}$=251~K.
    Planet c, with $M_p\sim$1.3$M_\oplus$, $T_{eq}$=400~K and also detectable by LIFE, is an interesting case study on the effect of stellar irradiation on rocky planets.
    
    \item[Teegarden's Star b and c:] To date, these planets are among the most similar to Earth in terms of mass, with $M_p\,sin\,i\sim$1.1$M_\oplus$ \citep{zechmeisteretal2019} and true masses that we estimate to be $\sim$1.27$M_\oplus$.
    They have almost circular orbits with periods of 4.9 and 11.4 days, and are likely tidally locked.
    The M7.0V star shows currently low activity \citep{zechmeisteretal2019}.
    \citet{wandel-talor2019} found that these planets are likely to sustain liquid water --at least on part of their surface-- for a wide range of atmospheric configurations.
    With no transits detected to date, LIFE will be the only mission able to probe the atmospheres of these targets.

    \item[Wolf 1061 c:] The three planets in the system were detected in RV \citep{wrightetal2016, astudillodefru2017}, with planet c being the only HZ one.
    We obtain for this planet a mass of about 4.1$M_\oplus$ and $T_{eq}$=261~K.
    \citet{wrightetal2016} estimated a transit probability for planet c of 5.9\%, but \citet{kaneetal2017} detected no transits in 7 years of ground-based photometry.
    LIFE will be capable of directly imaging this planet (Table \ref{table:detectability_onlyLIFE_20pc}) and will also detect planets b and d.
    
    \item[GJ 3323 b:] This system hosts two super-Earths with $M_p\,\sim\,2.3M_\oplus$ \citep{astudillodefru2017}, and we find GJ 3323 b (with $T_{eq}$=255~K) to orbit at the inner edge of the optimistic HZ.\footnote{\citet{astudillodefru2017} reported that planet c was the one in the HZ. We confirmed that this was due to an error in their Table 4 (Astudillo-Defru priv. comm.).} Only LIFE will be able to detect this planet, although the outer companion GJ 3323 c will remain out of reach because its $T_{eq}$=130~K will make it much fainter in the mid-IR.
    
    \item[GJ 667 C c, e and f:] GJ 667 C is an M1.5V star \citep{geballeetal2002} orbiting the pair of K stars GJ 667 AB at a distance of about 230~AU \citep[e.g.][]{angladaescudeetal2013}. 
    Planets b and c were the first detected in RV \citep{bonfilsetal2013}, with additional reported signals for up to five more planets (named b--h) \citep{angladaescudeetal2012, angladaescudeetal2013}.
    Subsequent reanalyses of the RV data confirmed planets b and c and found no evidence for the other planets \citep{feroz-hobson2014, robertson-mahadevan2014}.
    Currently the NASA Exoplanet Archive includes only planets b, c, e, f, and g, with the latter three marked as controversial.
    All the planets in the system have low eccentricities, and in particular planets c, e and f have all values around $e$=0.1.
    Only LIFE will be able to directly image the potential HZ planets in this system.
    
    \item[Ross 508 b:] This super-Earth orbits an M4.5 star located at 11~pc. Its orbit has a period of 10.7 days and a rather high eccentricity of 0.33 \citep{harakawaetal2022}. We estimate for Ross 508 b a mass of 4.8$M_\oplus$ and $T_{eq}$=266~K, placing it at the inner edge of the optimistic HZ. \citet{harakawaetal2022} found a transit probability of 1.6\% for this planet.

\end{description}

\begin{table*}     
\section{Young planetary systems} \label{sec:appendix_young}
\tiny
\centering
\caption{Detectability with LIFE of the known exoplanets that have been directly imaged up to date, as reported in the NASA Exoplanet Archive.}
\label{table:results_young_systems} 
\begin{tabular}{l c c c c c c c c c c}
\hline
\hline
Planet & $d$ & $a$ & $M_p$ & $R_p$ & \multicolumn{2}{c}{LIFE (4x1m)}  & \multicolumn{2}{c}{LIFE (4x2m)}   &  \multicolumn{2}{c}{LIFE (4x3.5m)} \\ 
 & [pc] & [AU] & [$M_J$] & [$R_J$] & $S/N$ & $t_{\rm{int}}$  & $S/N$ & $t_{\rm{int}}$  & $S/N$ & $t_{\rm{int}}$\\ 
 &	&	&	&	& (10h) & [h] & (10h) & [h] & (10h) & [h] \\ 
\hline
bet Pic b  &  19.75$^{ +0.09 }_{ -0.09 }$  &  11.30$^{ +2.10 }_{ -1.74 }$  &  $^{  }_{  }$  &  1.36$^{ +0.01 }_{ -0.01 }$  &  5498.9  &  2.0$\times 10^{-5}$ & 16024.0  &  0.0 & 33308.8 & 0.0  \\ 
HN Peg b  &  18.12$^{ +0.01 }_{ -0.01 }$  &  773.32$^{ +8.57 }_{ -8.90 }$  &  21.962$^{ +6.513 }_{ -6.305 }$  &  1.09$^{ +0.06 }_{ -0.07 }$  &  3029.8  &  5.0$\times 10^{-5}$ & 8892.8  &  1.0$\times 10^{-5}$ & 17268.8 & 0.0  \\ 
2MASS J02-39 b\tablefoottext{a}  &  39.40$^{ +0.00 }_{ -0.00 }$  &  156.37$^{ +6.30 }_{ -7.34 }$  &  13.853$^{ +0.790 }_{ -0.751 }$  &  1.44$^{ +0.02 }_{ -0.02 }$  &  2677.0  &  7.0$\times 10^{-5}$ & 8593.5  &  1.0$\times 10^{-5}$ & 18913.1 & 0.0  \\ 
HD 100546 b  &  109.69$^{ +0.42 }_{ -0.44 }$  &  52.96$^{ +1.41 }_{ -1.28 }$  &  $^{  }_{  }$  &  6.71$^{ +1.96 }_{ -1.76 }$  &  2521.6  &  8.0$\times 10^{-5}$ & 7369.5  &  1.0$\times 10^{-5}$ & 14678.0 & 0.0  \\ 
2MASS J21+16 b\tablefoottext{b}  &  33.10$^{ +0.00 }_{ -0.00 }$  &  3.95$^{ +0.47 }_{ -0.57 }$  &  52.666$^{ +35.262 }_{ -37.226 }$  &  0.94$^{ +0.24 }_{ -0.27 }$  &  2430.8  &  8.0$\times 10^{-5}$ & 7986.9  &  1.0$\times 10^{-5}$ & 17716.8 & 0.0  \\ 
Ross 458 c  &  11.51$^{ +0.01 }_{ -0.01 }$  &  1100.00$^{ +0.00 }_{ -0.00 }$  &  6.000$^{ +0.000 }_{ -0.000 }$  &  1.07$^{ +0.04 }_{ -0.05 }$  &  2372.3  &  9.0$\times 10^{-5}$ & 6108.8  &  1.0$\times 10^{-5}$ & 11774.5 & 0.0  \\ 
TYC 8998-760-1 b  &  94.62$^{ +0.20 }_{ -0.18 }$  &  162.00$^{ +0.00 }_{ -0.00 }$  &  13.900$^{ +2.109 }_{ -2.014 }$  &  2.73$^{ +0.32 }_{ -0.31 }$  &  2041.2  &  1.2$\times 10^{-4}$ & 6869.9  &  1.0$\times 10^{-5}$ & 14867.4 & 0.0  \\ 
AB Pic b  &  50.05$^{ +0.05 }_{ -0.05 }$  &  260.00$^{ +0.00 }_{ -0.00 }$  &  13.482$^{ +0.351 }_{ -0.344 }$  &  1.26$^{ +0.17 }_{ -0.16 }$  &  1962.4  &  1.3$\times 10^{-4}$ & 6840.7  &  1.0$\times 10^{-5}$ & 14793.0 & 0.0  \\ 
2MASS J01-24 b\tablefoottext{c}  &  33.83$^{ +0.06 }_{ -0.06 }$  &  51.87$^{ +4.28 }_{ -3.97 }$  &  24.513$^{ +1.736 }_{ -1.644 }$  &  0.99$^{ +0.15 }_{ -0.13 }$  &  1947.3  &  1.3$\times 10^{-4}$ & 6311.2  &  1.0$\times 10^{-5}$ & 14060.1 & 0.0  \\ 
2MASS J04+23 b\tablefoottext{d}  &  145.00$^{ +0.00 }_{ -0.00 }$  &  15.00$^{ +0.00 }_{ -0.00 }$  &  7.334$^{ +1.808 }_{ -1.594 }$  &  3.61$^{ +1.20 }_{ -1.36 }$  &  1801.5  &  1.5$\times 10^{-4}$ & 6065.7  &  1.0$\times 10^{-5}$ & 13210.6 & 0.0  \\ 
kap And b  &  50.06$^{ +0.56 }_{ -0.60 }$  &  55.04$^{ +1.30 }_{ -1.43 }$  &  24.513$^{ +8.812 }_{ -8.307 }$  &  1.25$^{ +0.10 }_{ -0.10 }$  &  1578.4  &  2.0$\times 10^{-4}$ & 5436.9  &  2.0$\times 10^{-5}$ & 12108.1 & 0.0  \\ 
ROXs 42 B b  &  143.60$^{ +1.07 }_{ -1.11 }$  &  157.00$^{ +0.00 }_{ -0.00 }$  &  8.914$^{ +2.077 }_{ -1.928 }$  &  2.49$^{ +0.15 }_{ -0.12 }$  &  1211.7  &  3.3$\times 10^{-4}$ & 4502.8  &  2.0$\times 10^{-5}$ & 10016.2 & 0.0  \\ 
PDS 70 b  &  113.08$^{ +0.33 }_{ -0.37 }$  &  20.06$^{ +1.31 }_{ -1.33 }$  &  2.967$^{ +0.686 }_{ -0.663 }$  &  2.88$^{ +0.22 }_{ -0.23 }$  &  1165.8  &  3.6$\times 10^{-4}$ & 4119.2  &  3.0$\times 10^{-5}$ & 8757.6 & 1.0$\times 10^{-5}$  \\ 
HIP 78530 b  &  136.70$^{ +1.03 }_{ -0.93 }$  &  741.60$^{ +39.77 }_{ -42.31 }$  &  22.987$^{ +0.647 }_{ -0.659 }$  &  1.84$^{ +0.10 }_{ -0.10 }$  &  1128.7  &  3.8$\times 10^{-4}$ & 4401.6  &  3.0$\times 10^{-5}$ & 10226.1 & 0.0  \\ 
HR 8799 c  &  41.24$^{ +0.11 }_{ -0.10 }$  &  38.00$^{ +0.00 }_{ -0.00 }$  &  9.952$^{ +2.058 }_{ -1.965 }$  &  1.00$^{ +0.14 }_{ -0.14 }$  &  1003.7  &  4.9$\times 10^{-4}$ & 3347.7  &  4.0$\times 10^{-5}$ & 6721.2 & 1.0$\times 10^{-5}$  \\ 
COCONUTS-2 b  &  10.89$^{ +0.00 }_{ -0.00 }$  &  9106.99$^{ +2375.80 }_{ -2510.44 }$  &  6.067$^{ +1.142 }_{ -1.149 }$  &  1.11$^{ +0.02 }_{ -0.02 }$  &  980.4  &  5.1$\times 10^{-4}$ & 2797.0  &  6.0$\times 10^{-5}$ & 5843.0 & 1.0$\times 10^{-5}$  \\ 
HR 8799 d  &  41.24$^{ +0.11 }_{ -0.11 }$  &  24.00$^{ +0.00 }_{ -0.00 }$  &  10.151$^{ +1.998 }_{ -2.196 }$  &  0.89$^{ +0.14 }_{ -0.13 }$  &  978.0  &  5.1$\times 10^{-4}$ & 3233.4  &  5.0$\times 10^{-5}$ & 6504.8 & 1.0$\times 10^{-5}$  \\ 
HD 106906 b  &  102.99$^{ +0.35 }_{ -0.29 }$  &  650.00$^{ +0.00 }_{ -0.00 }$  &  11.025$^{ +1.315 }_{ -1.421 }$  &  1.60$^{ +0.00 }_{ -0.00 }$  &  950.6  &  5.4$\times 10^{-4}$ & 3549.0  &  4.0$\times 10^{-5}$ & 7802.0 & 1.0$\times 10^{-5}$  \\ 
HR 8799 e  &  41.25$^{ +0.10 }_{ -0.10 }$  &  16.98$^{ +0.98 }_{ -1.11 }$  &  11.477$^{ +3.894 }_{ -3.677 }$  &  0.90$^{ +0.14 }_{ -0.14 }$  &  947.0  &  5.5$\times 10^{-4}$ & 3310.0  &  4.0$\times 10^{-5}$ & 6679.0 & 1.0$\times 10^{-5}$  \\ 
WISEP J12+16 b\tablefoottext{e}  &  10.10$^{ +0.00 }_{ -0.00 }$  &  8.07$^{ +0.87 }_{ -0.92 }$  &  21.917$^{ +1.418 }_{ -1.292 }$  &  0.87$^{ +0.02 }_{ -0.02 }$  &  920.1  &  5.8$\times 10^{-4}$ & 2743.1  &  7.0$\times 10^{-5}$ & 5735.7 & 1.0$\times 10^{-5}$  \\ 
GU Psc b  &  47.55$^{ +0.11 }_{ -0.11 }$  &  1999.45$^{ +135.25 }_{ -126.82 }$  &  11.368$^{ +1.131 }_{ -1.178 }$  &  1.18$^{ +0.02 }_{ -0.02 }$  &  871.6  &  6.4$\times 10^{-4}$ & 3124.1  &  5.0$\times 10^{-5}$ & 6630.2 & 1.0$\times 10^{-5}$  \\ 
PDS 70 c  &  113.06$^{ +0.34 }_{ -0.34 }$  &  35.42$^{ +3.26 }_{ -3.15 }$  &  1.998$^{ +0.702 }_{ -0.697 }$  &  2.44$^{ +0.40 }_{ -0.40 }$  &  856.5  &  6.7$\times 10^{-4}$ & 3025.0  &  5.0$\times 10^{-5}$ & 6415.4 & 1.0$\times 10^{-5}$  \\ 
CFHTWIR-Oph 98 b  &  137.00$^{ +0.00 }_{ -0.00 }$  &  199.91$^{ +4.19 }_{ -4.18 }$  &  7.732$^{ +0.541 }_{ -0.505 }$  &  1.86$^{ +0.03 }_{ -0.03 }$  &  832.8  &  7.1$\times 10^{-4}$ & 3160.4  &  5.0$\times 10^{-5}$ & 6978.2 & 1.0$\times 10^{-5}$  \\ 
HR 2562 b  &  34.01$^{ +0.03 }_{ -0.03 }$  &  20.30$^{ +0.20 }_{ -0.20 }$  &  29.554$^{ +10.914 }_{ -10.051 }$  &  0.56$^{ +0.01 }_{ -0.01 }$  &  739.4  &  9.0$\times 10^{-4}$ & 2471.8  &  8.0$\times 10^{-5}$ & 5066.1 & 2.0$\times 10^{-5}$  \\ 
1RXS J16-21 b\tablefoottext{f}  &  139.15$^{ +0.91 }_{ -0.89 }$  &  330.00$^{ +0.00 }_{ -0.00 }$  &  7.973$^{ +0.665 }_{ -0.645 }$  &  1.70$^{ +0.00 }_{ -0.00 }$  &  697.4  &  1.0$\times 10^{-3}$ & 2740.0  &  7.0$\times 10^{-5}$ & 6291.5 & 1.0$\times 10^{-5}$  \\ 
HIP 65426 b  &  108.90$^{ +0.49 }_{ -0.53 }$  &  92.00$^{ +0.00 }_{ -0.00 }$  &  9.220$^{ +1.905 }_{ -2.226 }$  &  1.50$^{ +0.02 }_{ -0.02 }$  &  679.3  &  1.1$\times 10^{-3}$ & 2616.7  &  7.0$\times 10^{-5}$ & 5730.7 & 1.0$\times 10^{-5}$  \\ 
2MASS J12-39 b\tablefoottext{g}  &  64.32$^{ +0.44 }_{ -0.45 }$  &  55.00$^{ +0.00 }_{ -0.00 }$  &  4.878$^{ +1.461 }_{ -1.333 }$  &  1.00$^{ +0.21 }_{ -0.21 }$  &  627.9  &  1.2$\times 10^{-3}$ & 2529.6  &  8.0$\times 10^{-5}$ & 5553.6 & 2.0$\times 10^{-5}$  \\ 
GJ 504 b  &  17.53$^{ +0.05 }_{ -0.05 }$  &  43.50$^{ +0.00 }_{ -0.00 }$  &  5.561$^{ +1.987 }_{ -1.786 }$  &  0.96$^{ +0.05 }_{ -0.05 }$  &  598.6  &  1.4$\times 10^{-3}$ & 1635.3  &  1.8$\times 10^{-4}$ & 3314.7 & 4.0$\times 10^{-5}$  \\ 
TYC 8998-760-1 c  &  94.63$^{ +0.18 }_{ -0.20 }$  &  320.00$^{ +0.00 }_{ -0.00 }$  &  6.007$^{ +0.679 }_{ -0.672 }$  &  1.25$^{ +0.31 }_{ -0.31 }$  &  489.2  &  2.1$\times 10^{-3}$ & 1998.6  &  1.2$\times 10^{-4}$ & 4297.1 & 3.0$\times 10^{-5}$  \\ 
HR 8799 b  &  41.25$^{ +0.10 }_{ -0.10 }$  &  68.00$^{ +0.00 }_{ -0.00 }$  &  7.952$^{ +2.030 }_{ -2.053 }$  &  0.60$^{ +0.07 }_{ -0.07 }$  &  469.9  &  2.2$\times 10^{-3}$ & 1624.9  &  1.9$\times 10^{-4}$ & 3293.2 & 5.0$\times 10^{-5}$  \\ 
51 Eri b  &  29.76$^{ +0.08 }_{ -0.08 }$  &  13.20$^{ +0.14 }_{ -0.14 }$  &  2.000$^{ +0.000 }_{ -0.000 }$  &  0.68$^{ +0.09 }_{ -0.10 }$  &  256.7  &  7.4$\times 10^{-3}$ & 767.9  &  8.3$\times 10^{-4}$ & 1493.2 & 2.2$\times 10^{-4}$  \\ 
HD 95086 b  &  86.23$^{ +0.15 }_{ -0.16 }$  &  55.77$^{ +1.68 }_{ -1.72 }$  &  4.987$^{ +1.340 }_{ -1.368 }$  &  0.85$^{ +0.07 }_{ -0.07 }$  &  220.8  &  1.0$\times 10^{-2}$ & 915.9  &  5.8$\times 10^{-4}$ & 1858.5 & 1.4$\times 10^{-4}$  \\ 
GQ Lup b  &  151.17$^{ +0.74 }_{ -0.72 }$  &  100.00$^{ +0.00 }_{ -0.00 }$  &  20.000$^{ +0.000 }_{ -0.000 }$  &  0.68$^{ +0.15 }_{ -0.14 }$  &  143.4  &  2.4$\times 10^{-2}$ & 838.0  &  7.0$\times 10^{-4}$ & 1853.4 & 1.4$\times 10^{-4}$  \\ 
\hline
\multicolumn{11}{l}{Notes. \parbox[t]{17 cm}{The values given here are the result of our statistical methodolody to simulate the planetary orbits (Sect. \ref{subsec:methods_orbits}). Planets for which the quoted parameter uncertainties are 0.00 generally lack a value for the uncertainties in the NASA Archive.}}
\end{tabular}
\tablefoottext{a}{2MASS J02192210-3925225 b}
\tablefoottext{b}{2MASS J21402931+1625183 A b}
\tablefoottext{c}{2MASS J01225093-2439505 b}
\tablefoottext{d}{2MASS J04414489+2301513 b}
\tablefoottext{e}{WISEP J121756.91+162640.2 A b}
\tablefoottext{f}{1RXS J160929.1-210524 b}
\tablefoottext{g}{2MASS J12073346-3932539 b}
\end{table*}
\FloatBarrier
We compiled the list of young planets that have been already directly imaged from ground-based or space-borne observatories.
We used the NASA Exoplanet Archive as reference and added the values of effective temperatures and radii from the literature. Table \ref{table:results_young_systems} shows the LIFE integration times for these planets. In all cases we assumed 3 exozodis, although we note that some of these systems will have higher exozodi levels. We did not compute the observability with HWO since most of these young planetary systems were discovered using direct imaging techniques in the near-infrared (J, H, K-bands). Virtually all of these planets should also be detectable with future reflected-light missions in the visible (unless the OWA becomes a limitation), given that they will achieve much deeper contrasts compared to ground-based facilities.

The vast majority of these directly-imaged planets will already have their mid-IR spectra being measured with \emph{JWST} or the ground-based ELTs and will not be prime targets for LIFE. However,  constraining the required integration times with LIFE enables an interesting comparison between the different facilities. We note that \emph{JWST} has roughly the same collecting power as the optimistic scenario of LIFE with four 3.5~m mirrors.
\FloatBarrier

        
\onecolumn


\begin{landscape}
\section{Complete output catalogue} \label{sec:appendix_longtables}
\small
\begin{longtable}{ l  c  c  c  c  c  c  c  c  c  c  c  c  c } 
\caption{\label{table:output_catalogue} 
Catalogue of all the known exoplanets within 20 pc produced with our statistical methodology (see Sect. \ref{sec:methods}). We report, for each parameter, the median values of the probability distributions resulting from the 1000 orbital realizations and the uncertainties given by the 16\% and 84\% percentiles.}\\
 \hline 
 \hline
  Planet & {$d$} & {$P$} & {$a$} & {$M_p$} & {$R_p$} & {$i$} & $e$ & {$\omega_p$} & $T_{eq(mean)}$ & $R_\star$ & $T_\star$ & $V$ & St. type \\
  & [pc] & [days] & [AU] & [$M_{J}$] & [$R_{J}$] & [deg] & & [deg] & [K] & [$R_\odot$] & [K] & [mag] & \\
  \hline
  \endfirsthead
\caption{continued.}\\
\hline
\hline                
  Planet & {$d$} & {$P$} & {$a$} & {$M_p$} & {$R_p$} & {$i$} & $e$ & {$\omega_p$} & $T_{eq(mean)}$ & $R_\star$ & $T_\star$ & $V$ & St. type \\
  & [pc] & [days] & [AU] & [$M_{J}$] & [$R_{J}$] & [deg] & & [deg] & [K] & [$R_\odot$] & [K] & [mag] & \\
\hline
\endhead
\hline
\endfoot
\tiny
Proxima Cen b & 1.30$^{ +0.00 }_{ -0.00 }$ & 11.2$^{ +0.0 }_{ -0.0 }$ & 0.05$^{ +0.00 }_{ -0.00 }$ & 0.005$^{ +0.003 }_{ -0.001 }$ & 0.11$^{ +0.01 }_{ -0.01 }$ & 90.1$^{ +42.9 }_{ -45.5 }$ & 0.35$^{ +0.00 }_{ -0.00 }$ & 130.0$^{ +0.0 }_{ -0.0 }$ & 229$^{ +34 }_{ -37 }$ & 0.14$^{ +0.02 }_{ -0.02 }$ & 3050$^{ +100 }_{ -100 }$ & 11.0 & M5.5 V \\ 
eps Eri b & 3.20$^{ +0.00 }_{ -0.00 }$ & 2667.7$^{ +13.8 }_{ -13.2 }$ & 3.53$^{ +0.02 }_{ -0.02 }$ & 0.676$^{ +0.069 }_{ -0.072 }$ & 1.22$^{ +0.02 }_{ -0.02 }$ & 83.3$^{ +16.5 }_{ -17.8 }$ & 0.08$^{ +0.04 }_{ -0.04 }$ & 162.2$^{ +58.3 }_{ -65.0 }$ & 97$^{ +3 }_{ -3 }$ & 0.76$^{  }_{  }$ & 5020$^{  }_{  }$ & 3.7 & K2 V \\ 
GJ 887 b & 3.29$^{ +0.01 }_{ -0.01 }$ & 9.3$^{ +0.0 }_{ -0.0 }$ & 0.07$^{ +0.00 }_{ -0.00 }$ & 0.016$^{ +0.009 }_{ -0.003 }$ & 0.18$^{ +0.06 }_{ -0.03 }$ & 89.3$^{ +44.5 }_{ -42.8 }$ & 0.00$^{ +0.00 }_{ -0.00 }$ & 180.9$^{ +116.8 }_{ -120.8 }$ & 404$^{ +24 }_{ -28 }$ & 0.47$^{ +0.09 }_{ -0.09 }$ & 3688$^{ +86 }_{ -86 }$ & 7.6 & M1 V \\ 
GJ 887 c & 3.29$^{ +0.01 }_{ -0.01 }$ & 21.8$^{ +0.0 }_{ -0.0 }$ & 0.12$^{ +0.00 }_{ -0.00 }$ & 0.028$^{ +0.017 }_{ -0.004 }$ & 0.26$^{ +0.08 }_{ -0.04 }$ & 89.2$^{ +43.2 }_{ -42.4 }$ & 0.48$^{ +0.37 }_{ -0.33 }$ & 180.4$^{ +117.4 }_{ -124.0 }$ & 331$^{ +209 }_{ -55 }$ & 0.47$^{ +0.09 }_{ -0.09 }$ & 3688$^{ +86 }_{ -86 }$ & 7.6 & M1 V \\ 
Ross 128 b & 3.37$^{ +0.00 }_{ -0.00 }$ & 9.9$^{ +0.0 }_{ -0.0 }$ & 0.05$^{ +0.00 }_{ -0.00 }$ & 0.005$^{ +0.003 }_{ -0.001 }$ & 0.11$^{ +0.01 }_{ -0.01 }$ & 89.1$^{ +41.8 }_{ -42.2 }$ & 0.12$^{ +0.07 }_{ -0.07 }$ & 177.6$^{ +124.6 }_{ -120.2 }$ & 267$^{ +16 }_{ -13 }$ & 0.20$^{ +0.01 }_{ -0.01 }$ & 3192$^{ +60 }_{ -60 }$ & 11.1 & M4 \\
\hline
\end{longtable}
\tablefoot{
In those cases where the quoted uncertainties are 0.00, this might be due to a lack of reported uncertainties in the NASA Archive or a result of insufficient significant figures in the rounding.
}
\tablefoottext{a}{VHS J125601.92-125723.9 b}
\tablefoottext{b}{TYC 2187-512-1 b}
\tablefoottext{c}{LSPM J2116+0234 b}
\end{landscape}

\end{appendix}

\end{document}